\documentclass[12pt]{article}
\pdfoutput=1

\usepackage[a4paper,text={16.8cm,22.4cm}]{geometry}
\usepackage{amsmath,amsfonts,braket,slashed,amssymb,bm,psfrag,graphicx,color,colortbl,dsfont,euscript,ctable,bbm}
\usepackage[small,labelfont=bf]{caption}
\RequirePackage[sort&compress,square,comma,numbers]{natbib}
\allowdisplaybreaks
\newcommand{\spac}{{\hspace{0.3mm}}}
\newcommand{\negspac}{{\hspace{-0.4mm}}}
\newcommand{\iDslLR}{i\hspace{-0.9mm}\overleftrightarrow{\rlap{\hspace{0.8mm}/}{D}}}

\newcommand{\blueterms}[1]{{\textcolor{black}{#1}}}

\definecolor{LightCyan}{rgb}{0.88,1,1}
\definecolor{LightGray}{rgb}{0.85,0.85,0.85}

\begin{document}

\begin{titlepage}

\begin{flushright}
\normalsize
MITP/21-021\\
SISSA 10/2021/FISI\\
ZU-TH-18/21\\
May 3, 2021
\end{flushright}

\vspace{3mm}
\begin{center}
\huge\bf
ALP\,\negspac--\,SMEFT Interference
\end{center}

\vspace{-1mm}
\begin{center}
Anne Mareike Galda$^a$, Matthias Neubert$^{a,b,c}$ and Sophie Renner$^d$\\
\vspace{0.7cm} 
{\sl ${}^a$PRISMA$^+$ Cluster of Excellence \& Mainz Institute for Theoretical Physics\\
Johannes Gutenberg University, 55099 Mainz, Germany\\[2mm]
${}^b$Physik-Institut, Universit\"at Z\"urich, CH-8057, Switzerland\\[2mm]
${}^c$Department of Physics \& LEPP, Cornell University, Ithaca, NY 14853, U.S.A.\\[2mm]
${}^d$SISSA International School for Advanced Studies \& INFN, Sezione di Trieste\\
Via Bonomea 265, 34136, Trieste, Italy}
\end{center}

\vspace{0.8cm}
\begin{abstract}
The Standard Model Effective Field Theory (SMEFT) offers a powerful theoretical framework for parameterizing the low-energy effects of heavy new particles with masses far above the scale of electroweak symmetry breaking. Additional light degrees of freedom extend the effective theory. We show that light new particles that are weakly coupled to the SM via non-renormalizable interactions induce non-zero Wilson coefficients in the SMEFT Lagrangian via renormalization-group evolution. For the well-motivated example of axions and axion-like particles (ALPs) interacting with the SM via classically shift-invariant dimension-5 interactions, we calculate how these interactions contribute to the one-loop renormalization of the dimension-6 SMEFT operators, and how this running sources additional contributions to the Wilson coefficients on top of those expected from heavy new states. As an application, we study the ALP contributions to the magnetic dipole moment of the top quark and comment on implications of electroweak precision constraints on ALP couplings.
\end{abstract}

\end{titlepage}

\tableofcontents

\section{Introduction}

The absence of a discovery of new particles at the LHC suggests that additional degrees of freedom not contained in the Standard Model (SM) of particle physics have masses far above the scale of electroweak symmetry breaking. Existing hints of departures from the SM, indicated for instance by recent data on  semileptonic decays of $B$ mesons \cite{Aaij:2021vac} and the anomalous magnetic moment of the muon \cite{Abi:2021gix}, are in line with this expectation. Assuming that the theory above the electroweak scale respects the full SM gauge symmetry and that the Higgs is an $SU(2)_L$ doublet, the Standard Model Effective Field Theory (SMEFT) \cite{Buchmuller:1985jz} provides a systematic approach for describing the virtual effects of heavy new particles on low-energy precision measurements (for a review, see \cite{Brivio:2017vri}).

On the other hand, there remains the possibility of the existence of light new particles interacting very weakly with the SM, for instance because these interactions are mediated by higher-dimensional operators. A prominent example are axions and axion-like particles, to which we will collectively refer as ALPs in this work. The existence of such particles is theoretically well motivated because of their potential relevance for solving the strong CP problem \cite{Peccei:1977hh,Weinberg:1977ma,Wilczek:1977pj}. More generally, ALPs are gauge-singlet pseudoscalar particles, which arise in many theories beyond the SM as pseudo Nambu--Goldstone bosons of a global symmetry spontaneously broken in the ultra-violet (UV) theory, and are thus naturally lighter than other new states whose masses may be out of reach of current direct searches. While the QCD axion is the original example of such a particle, ALPs with a variety of masses and couplings can arise also more generally, for example in theories which address the flavor \cite{Calibbi:2016hwq,Ema:2016ops} and hierarchy \cite{Bagger:1994hh,Gripaios:2009pe,Ferretti:2013kya,Graham:2015cka,Bellazzini:2017neg} problems. Here we remain agnostic about the underlying theory and study a general effective field theory, to which explicit models can be matched. Since the ALP couplings to SM fields are constrained by a shift symmetry, these interactions first appear at the level of dimension-5 operators in the effective theory.  

The existence of a light ALP extends the SMEFT, because operators can now be built out of SM fields and the ALP field. The most general effective Lagrangian above the electroweak scale can then be written in the form
\begin{equation}\label{completeL}
   {\cal L}_{\rm eff} 
   = {\cal L}_{\rm SM} + \frac12 \left( \partial_\mu a\right)\!\left( \partial^\mu a\right) 
    - \frac{m_a^2}{2}\,a^2 + {\cal L}_{\rm SM+ALP} + {\cal L}_{\rm SMEFT} \,, 
\end{equation}
where the last two terms consist of higher-dimensional operators starting at dimension-5 order. By definition the term ${\cal L}_{\rm SMEFT}$ consists of operators {\em not\/} containing the ALP field. We will refer to these operators as ``SMEFT operators'', although of course the theory we are working in is not the pure SMEFT, but rather the effective theory whose degrees of freedom are the SM fields plus the ALP. Like any other new particle not contained in the SM, an ALP can be searched for either directly or indirectly. Direct searches using particle-physics detectors aim at detecting ALPs produced in a collider experiment by reconstructing them in various possible decay modes \cite{Chen:2010su,Mimasu:2014nea,Jaeckel:2015jla,Knapen:2016moh,Brivio:2017ije,Bauer:2017nlg,Bauer:2017ris,Mariotti:2017vtv,Craig:2018kne,Bauer:2018uxu,Alonso-Alvarez:2018irt,Gavela:2019cmq}. Our focus in this work is on indirect searches, which look for hints of the effects of virtual ALP exchange on precision measurements. As long as the scale $f$ suppressing the ALP couplings to the SM is not excessively large,\footnote{Otherwise one needs to resort to cosmological and astrophysical bounds, see e.g.\
\cite{Cadamuro:2011fd,Millea:2015qra,Payez:2014xsa,Jaeckel:2017tud}.} 
this opens up new, complementary ways to search for the effects of an ALP and place bounds on its couplings to the SM. An advantage is that, contrary to direct searches, indirect probes are independent of the way in which the ALP decays, and if it is long- or short-lived. Prominent examples of such precision observables are the anomalous magnetic moment of the muon and the electron \cite{Marciano:2016yhf,Bauer:2017nlg,Bauer:2017ris,Bauer:2019gfk,Buen-Abad:2021fwq}, electroweak precision observables \cite{Bauer:2017ris}, and various rates for flavor-changing decays of kaons, $B$-mesons and charged leptons \cite{Dolan:2014ska,Izaguirre:2016dfi,Choi:2017gpf,Arias-Aragon:2017eww,Gavela:2019wzg,MartinCamalich:2020dfe,Chakraborty:2021wda,Bauer:2021wjo}. It has been observed in several of these studies that the one-loop contributions to such quantities arising from virtual ALP exchange can be ultra-violet (UV) divergent, but a consistent treatment of these divergences has so far not been presented in the literature.

The renormalization-group (RG) evolution of the dimension-5 ALP couplings in the effective Lagrangian ${\cal L}_{\rm SM+ALP}$ has recently been studied in \cite{Bauer:2020jbp,Chala:2020wvs}. Here we point out the important fact that the existence of an ALP with couplings to the SM {\em necessarily\/} induces non-zero Wilson coefficients of the dimension-6 SMEFT operators contained in ${\cal L}_{\rm SMEFT}$, which produce effects across a wide variety of observables. In general, there are three types of such effects: 
\begin{itemize}
\item 
Matching contributions at the scale $\Lambda=4\pi f$ of global symmetry breaking arise from integrating out heavy new particles contained in the UV completion of the ALP model. Examples are heavy vector-like fermions with different Peccei--Quinn charges, by which the ALP interacts with SM gauge bosons. These matching contributions are model dependent, and they can only be calculated within a specific UV completion.
\item
Loop diagrams involving virtual exchange of the ALP are generally divergent and induce inhomogeneous source terms in the RG equations for the Wilson coefficients of the SMEFT operators. 
\item 
At low energies, the time-ordered product with two insertions of the ALP effective Lagrangian ${\cal L}_{\rm SM+ALP}$ yields non-zero contributions to scattering amplitudes describing processes involving SM particles only. These contributions can be systematically calculated in the effective theory described by (\ref{completeL}) as long as the ALP mass or the characteristic scale of the observable are in the realm of perturbative QCD.
\end{itemize}
Of these three contributions, the second one is parametrically enhanced by large logarithms arising from the evolution from the high scale $\Lambda$ to low energies.\footnote{It is important to keep in mind that the other two contributions must also exist, if only to cancel the renormalization-scheme dependence of the second contribution.} 
For example, a contribution of this sort underlies ALP explanations for the deviation of the muon anomalous magnetic moment from its SM prediction \cite{Marciano:2016yhf,Bauer:2017ris,Buen-Abad:2021fwq}, for which divergent diagrams involving the ALP--photon coupling induce large logarithms in the coefficients of SMEFT dipole operators. 

In this work, we calculate for the first time the full set of ALP-induced terms in the RG equations for the Wilson coefficients of the dimension-6 SMEFT operators in (\ref{completeL}) above the electroweak scale. Irrespective of its mass, which can even be much smaller than the weak scale, the presence of an ALP generates inhomogeneous source terms in the RG equations, which we write in the form 
\begin{equation}\label{SMEFTrge}
   \frac{d}{d\ln\mu}\,C_i^{\rm SMEFT} - \gamma_{ji}^{\rm SMEFT}\,C_j^{\rm SMEFT} 
   = \frac{S_i}{(4\pi f)^2} \qquad \text{(for $\mu<4\pi f$)} \,.
\end{equation}
Here $\bm{\gamma}^{\rm SMEFT}$ is the anomalous-dimension matrix of the dimension-6 SMEFT operators in the Warsaw basis \cite{Grzadkowski:2010es} (the transpose matrix governs the evolution of the Wilson coefficients), which has been calculated at one-loop order in \cite{Elias-Miro:2013mua,Jenkins:2013zja,Jenkins:2013wua,Alonso:2013hga}. The ALP source terms are denoted by $S_i$, and the overall suppression scale is set by the ``ALP decay constant''  $f$. The presence of these source terms generates non-zero SMEFT Wilson coefficients irrespective of the existence of any other source of new physics. We find that almost the entire set of Wilson coefficients is sourced by ALP effects at one-loop order. As an important application of our results, we present a study of the chromo-magnetic moment of the top quark and briefly comment on constraints from electroweak precision observables. 

In our calculations we adopt the notations and conventions introduced in \cite{Grzadkowski:2010es}, with one exception: we define the covariant derivative in the fundamental representation of the gauge group as $D_\mu=\partial_\mu-ig_s\spac G_\mu^a\, T^a-ig_2 W_\mu^I\spac\frac{\sigma^I}{2}-ig_1\spac{\cal Y} B_\mu$, where $T^a$ are the Gell-Mann matrices, $\sigma^I$ the Pauli matrices and ${\cal Y}$ the hypercharge generator. While this agrees with most textbooks on quantum field theory, it corresponds to a different sign convention for the three gauge couplings compared with \cite{Grzadkowski:2010es,Alonso:2013hga}. While the Warsaw basis defines the standard choice of the SMEFT operators, it is convenient for many applications to redefine the operators containing field-strength tensors by absorbing powers of the gauge couplings into the operators (see e.g.\ \cite{Elias-Miro:2013mua}). In this way, odd powers of the gauge couplings $g_i$ and mixed terms involving $g_i g_j$ with $i\ne j$ can be avoided in the RG equations (\ref{SMEFTrge}). In Appendix~\ref{app:A} we introduce such a basis of redefined operators $\{\spac Q_i'\spac\}$ and present our results for the corresponding source terms $\{\spac S_i'\spac\}$.

In Section~\ref{sec:2} we briefly review the effective Lagrangian describing the interactions of ALPs with SM particles at dimension-5 order, following closely the presentation in \cite{Bauer:2020jbp}. In Section~\ref{sec:3} we explain our method for calculating the ALP contributions to the RG equations of the SMEFT operators. Our results for the complete list of ALP source terms $S_i$ in (\ref{SMEFTrge}), obtained at one-loop order in perturbation theory, are reported in Section~\ref{sec:4}. In Section~\ref{sec:applications} we briefly discuss two applications of our results, before concluding in Section~\ref{sec:conclusions}.

\section{ALP couplings to the SM}
\label{sec:2}

An ALP provides a paradigmatic example of a new light particle, which interacts weakly with the SM via effective interactions suppressed by a large scale $f\gg m_a$. Concretely, we consider a gauge-singlet, pseudoscalar resonance $a$, whose couplings to SM fields are protected by an approximate (classical) shift symmetry, possibly broken softly by the presence of an explicit mass term $m_a$ in (\ref{completeL}). This mass parameter is absent for the QCD axion. At dimension-5 order, the most general effective Lagrangian for such a particle can be written in the form \cite{Georgi:1986df} 
\begin{equation}\label{Leff}
\begin{aligned}
   {\cal L}_{\rm SM+ALP}^{D=5}
   &= c_{GG}\,\frac{\alpha_s}{4\pi}\,\frac{a}{f}\,G_{\mu\nu}^a\,\tilde G^{\mu\nu,a}
    + c_{WW}\,\frac{\alpha_2}{4\pi}\,\frac{a}{f}\,W_{\mu\nu}^I\,\tilde W^{\mu\nu,I}
    + c_{BB}\,\frac{\alpha_1}{4\pi}\,\frac{a}{f}\,B_{\mu\nu}\,\tilde B^{\mu\nu} \\  
   &\quad+ \frac{\partial^\mu a}{f}\,\sum_F\,\bar\psi_F\spac\bm{c}_F\spac\gamma_\mu\spac\psi_F \,.
\end{aligned}
\end{equation}
Here $G_{\mu\nu}^a$, $W_{\mu\nu}^I$ and $B_{\mu\nu}$ are the field-strength tensors of $SU(3)_c$, $SU(2)_L$ and $U(1)_Y$, while $\alpha_s=g_s^2/(4\pi)$, $\alpha_2=g_2^2/(4\pi)$ and $\alpha_1=g_1^2/(4\pi)$ denote the corresponding coupling parameters. $\tilde B^{\mu\nu}=\frac12\epsilon^{\mu\nu\alpha\beta} B_{\alpha\beta}$ etc.\ (with $\epsilon^{0123}=1$) are the dual field-strength tensors. The sum in the last term extends over the chiral fermion multiplets of the SM ($F=Q, L, u, d, e$). The quantities $\bm{c}_F$ are hermitian matrices in generation space. For the couplings of $a$ to the $SU(2)_L$ and $U(1)_Y$ gauge fields, the additional terms arising from a constant shift $a\to a+c$ of the ALP field can be removed by field redefinitions. For the coupling to gluons, instead, the continuous shift symmetry is broken by instanton effects to the discrete subgroup $a\to a+n\spac\pi f/c_{GG}$ with integer $n$ \cite{Weinberg:1977ma,Wilczek:1977pj}. In this process, the ALP receives a small mass even if the Lagrangian parameter $m_a$ in (\ref{completeL}) vanishes \cite{Bardeen:1978nq,Shifman:1979if,DiVecchia:1980yfw}.

Above we have indicated the suppression of the dimension-5 operators with the ALP decay constant $f$, which is related to the relevant new-physics scale by $\Lambda=4\pi f$. This is the characteristic scale of global (Peccei--Quinn) symmetry breaking, which we assume to be far above the electroweak scale. Higher-order interactions of ALPs with SM particles also exist, an important example being the dimension-6 operator $(\partial_\mu a)(\partial^\mu a)\spac H^\dagger H$, which mediates a coupling to the Higgs boson. However, this operator is suppressed by $1/f^2$, and hence it can only have an impact on the RG evolution of SMEFT operators of dimension~7 and higher.

Since our effective theory respects the SM gauge invariance and only contains the SM particles and the ALP as degrees of freedom, it would need to be modified in scenarios with a new-physics sector between the electroweak scale and the scale of global symmetry breaking ($v<M_{\rm NP}<4\pi f$). Even in this case, the effective Lagrangian (\ref{Leff}) offers a model-independent description of the physics below the intermediate scale $M_{\rm NP}$. 

Note the important fact that the effective Lagrangian (\ref{Leff}) does not contain a coupling of the ALP to the Higgs field. The renormalizable portal interaction $a^2\spac H^\dagger H$ is forbidden by the shift symmetry, while a possible shift-symmetric dimension-5 coupling of $\partial^\mu a$ to the Higgs current is redundant and can be removed by field redefinitions \cite{Georgi:1986df}. The free parameters of the model are the ALP mass $m_a$, the three ALP couplings $c_{VV}$ to gauge fields (with $V=G,W,B$), and the 5 times 9 parameters of the hermitian matrices $\bm{c}_F$.\footnote{The scale $f$ can be absorbed into the ALP couplings and hence does not add a new parameter.} 
It is well known that the derivative ALP couplings to fermions are only defined modulo the generators of exact global symmetries of the SM \cite{Georgi:1986df}, which are baryon number $B$ and the lepton numbers $L_e$, $L_\mu$ and $L_\tau$ for each flavor (since the neutrinos are massless in the SM). It follows that four model parameters are redundant and can be chosen at will. For example, one can choose $\left(\bm{c}_L\right)_{ii}=0$ for $i=1,2,3$ and $\text{Tr}(\bm{c}_Q)=0$, or one can arrange that either $c_{WW}=0$ or $c_{BB}=0$ (but not both) in addition with three constraints imposed on the ALP--fermion couplings. The model thus contains $1+3+45-4=45$ free parameters, most of them related to the flavor sector. 

The form of the effective Lagrangian (\ref{Leff}) is the one in which the shift symmetry is most explicit. However, for our purposes it will be useful to consider an alternative but equivalent form, in which the ALP couplings to fermions are of a non-derivative type \cite{Bauer:2020jbp}. Integrating by parts in the last term in (\ref{Leff}) and using the SM equations of motion (EOMs) along with the equation for the axial anomaly leads to (with $\tilde H_i=\epsilon_{ij}\spac H_j^*$) 
\begin{equation}\label{Leffalt}
\begin{aligned}
   {\cal L}_{\rm SM+ALP}^{D=5\,\prime}
   &= C_{GG}\,\frac{a}{f}\,G_{\mu\nu}^a\,\tilde G^{\mu\nu,a}
    + C_{WW}\,\frac{a}{f}\,W_{\mu\nu}^I\,\tilde W^{\mu\nu,I}
    + C_{BB}\,\frac{a}{f}\,B_{\mu\nu}\,\tilde B^{\mu\nu} \\
   &\quad\mbox{}- \frac{a}{f} \left( \bar Q\spac\tilde H\spac\widetilde{\bm{Y}}_u\spac u_R 
    + \bar Q\spac H\spac\widetilde{\bm{Y}}_d\,d_R 
    + \bar L\spac H\spac\widetilde{\bm{Y}}_e\spac e_R + \text{h.c.} \right) ,
\end{aligned}
\end{equation}
where the three Yukawa-type matrices $\widetilde{\bm{Y}}_f$ (with $f=u,d,e$) are related to the SM Yukawa matrices and the five hermitian matrices $\bm{c}_F$ by
\begin{equation}\label{rela1}
   \widetilde{\bm{Y}}_u = i\hspace{0.3mm} \big( \bm{Y}_u\,\bm{c}_u - \bm{c}_Q \bm{Y}_u \big) \,, \qquad
   \widetilde{\bm{Y}}_d = i\hspace{0.3mm} \big( \bm{Y}_d\,\bm{c}_d - \bm{c}_Q \bm{Y}_d \big) \,, \qquad
   \widetilde{\bm{Y}}_e = i\hspace{0.3mm} \big( \bm{Y}_e\,\bm{c}_e - \bm{c}_L \bm{Y}_e \big) \,.
\end{equation}
Note the important fact that the ALP--boson couplings in (\ref{Leff}) are also affected by the field redefinitions. One finds
\begin{equation}\label{eq:6}
\begin{aligned}
   C_{GG} &= \frac{\alpha_s}{4\pi} \left[
    c_{GG} + \frac12\,\text{Tr} \left( \bm{c}_d + \bm{c}_u - 2\spac\bm{c}_Q \right) \right] 
    \equiv \frac{\alpha_s}{4\pi}\,\tilde c_{GG} \,, \\
   C_{WW} &= \frac{\alpha_2}{4\pi} \left[
    c_{WW} - \frac12\,\text{Tr} \left( N_c\spac\bm{c}_Q + \bm{c}_L \right) \right] 
    \equiv \frac{\alpha_2}{4\pi}\,\tilde c_{WW} \,, \\
   C_{BB} &= \frac{\alpha_1}{4\pi}\,\bigg[ 
    c_{BB} + \text{Tr}\,\Big[ N_c \left( {\cal Y}_d^2\,\bm{c}_d 
    + {\cal Y}_u^2\,\bm{c}_u - 2\,{\cal Y}_Q^2\,\bm{c}_Q \right)
    + {\cal Y}_e^2\,\bm{c}_e - 2\,{\cal Y}_L^2\,\bm{c}_L \Big] \bigg] 
    \equiv \frac{\alpha_1}{4\pi}\,\tilde c_{BB} \,,
\end{aligned}
\end{equation}
where the traces are over generation indices, $N_c=3$ is the number of colors, and ${\cal Y}_u=\frac23$, ${\cal Y}_d=-\frac13$, ${\cal Y}_Q=\frac16$, ${\cal Y}_e=-1$ and ${\cal Y}_L=-\frac12$ denote the hypercharge quantum numbers of the SM quarks and leptons. Note that the couplings $C_{VV}$ and $\widetilde{\bm{Y}}_f$ in the effective Lagrangian (\ref{Leffalt}) are invariant under the redundancies mentioned above, i.e.\ the values of these parameters do not change under the field redefinitions corresponding to the generators of $B$, $L_e$, $L_\mu$ and $L_\tau$.

If the Lagrangian (\ref{Leffalt}) were taken as the defining effective Lagrangian of the ALP model, there would be no apparent reason for the complex matrices $\widetilde{\bm{Y}}_f$ to have any particular structure. The model would then be characterized by the ALP mass parameter $m_a$, the three ALP--boson couplings $C_{VV}$, and three generic complex matrices containing 3 times 18 parameters, making for a total of $1+3+54=58$ free parameters, 13 more than in the effective Lagrangian (\ref{Leff}). It is the shift symmetry encoded in the effective ALP Lagrangian (\ref{Leff}) that gives rise to the hierarchical structure of the matrices $\widetilde{\bm{Y}}_f$, which results from the appearance of the SM Yukawa matrices in (\ref{rela1}). This feature distinguishes an ALP from a generic pseudoscalar boson $a$. 

Our definitions of the ALP couplings in (\ref{Leff}) are such that the parameters $c_{VV}$ and $\bm{c}_F$ are expected to be of ${\cal O}(1)$ when one applies the counting rules of naive dimensional analysis \cite{Manohar:1983md,Luty:1997fk,Cohen:1997rt}. These rules imply, in particular, that the ALP--boson couplings $c_{VV}$ should be accompanied by a loop factor \cite{Bauer:2017ris}. This is natural in QCD axion models, because the $\theta$ parameter, which is dynamically set to zero by a shift of the axion field in the Peccei--Quinn mechanism~\cite{Peccei:1977hh}, couples to the topological quantity $\frac{\alpha_s}{8\pi}\,G_{\mu\nu}^a\spac\tilde G^{\mu\nu,a}$. In the literature on ALPs, on the other hand, the couplings to gauge bosons are often defined as in (\ref{Leffalt}), without factoring out one-loop suppression factors, and the parameters $C_{VV}$ are treated as phenomenological parameters. In our analysis we will be agnostic about the relative size of the various coupling parameters in (\ref{Leffalt}) and perform all calculations consistently to one-loop order in the ALP interactions. This is in the spirit of the Warsaw basis, where one avoids including loop factors in the definitions of the operators and making assumptions about the relative size of Wilson coefficients.

In recent work, a detailed study of the RG evolution of the ALP couplings to SM particles in the effective ALP Lagrangians (\ref{Leff}) and (\ref{Leffalt}) from the new-physics scale $\Lambda$ down to low energies has been performed \cite{Bauer:2020jbp,Chala:2020wvs}. We summarize the one-loop RG equations in Appendix~\ref{app:B}. In addition to the effects considered in these works, loop diagrams containing ALPs as virtual particles require dimension-6 (and higher) operators built out of SM fields as counterterms. The presence of an ALP, even if it is very light, thus generates non-zero Wilson coefficients of many of the operators in the SMEFT. In this paper we present a systematic study of these effects at one-loop order.

\section{Green's functions requiring SMEFT counterterms}
\label{sec:3}

In order to calculate the ALP source terms in (\ref{SMEFTrge}) from the effective Lagrangian (\ref{Leffalt}), we study the UV divergences of all relevant amputated, connected Green's functions involving a virtual ALP exchange, whose counterterms correspond to local dimension-6 SMEFT operators. We work consistently at one-loop order and in a general covariant gauge. All results shown below are gauge independent. To confine the analysis to one-particle irreducible (1PI) diagrams, we work with a basis of SMEFT operators that is complete without using the EOMs for the SM fields (this is sometimes called a Green's basis \cite{Gherardi:2020det}). After identifying the relevant counterterms in this basis, we project the results onto the Warsaw basis by eliminating the redundant operators. Following \cite{Grzadkowski:2010es}, we differentiate between purely bosonic operators, operators containing a single fermion current, and four-fermion operators. Table~\ref{tab:survey} summarizes which of the operators in the extended basis are contained in the Warsaw basis, and which of these operators are generated by ALP exchange either directly (via one-loop diagrams) or indirectly after using the EOMs. In each case we refer to the relevant figure for some representative Feynman diagrams. 

\begin{table}[p]
\centering
\begin{tabular}{ccccc}
\toprule
Operator class & Warsaw basis & \multicolumn{2}{c}{Way of generation} & Feynman graphs \\ 
\midrule
\rowcolor{LightGray}
Purely bosonic & & & & \\
\midrule
$X^3$ & yes & direct & --- & Figure~\ref{fig:X3} \\
\rowcolor{LightCyan}
$X^2 D^2$ & no & direct & & Figure~\ref{fig:X3} \\
$X^2 H^2$ & yes & direct & --- & Figure~\ref{fig:X2H2} \\
\rowcolor{LightCyan}
$X H^2 D^2$ & no & --- & & \\
$H^6$ & yes & --- & EOM & Figure~\ref{fig:X3} \\
$H^4 D^2$ & yes & --- & EOM & Figure~\ref{fig:X3} \\
\rowcolor{LightCyan}
$H^2 D^4$ & no & --- & & \\
\midrule
\rowcolor{LightGray}
Single fermion current & & & & \\
\midrule
\rowcolor{LightCyan}
$\psi^2 X D$ & no & --- & & \\
\rowcolor{LightCyan}
$\psi^2 D^3$ & no & --- & & \\
$\psi^2 X H$ & yes & direct & --- & Figure~\ref{fig:psi2XH} \\
$\psi^2 H^3$ & yes & direct & EOM & Figs.~\ref{fig:psi2H3},\! \ref{fig:X3},\! \ref{fig:psi2H2D} \\
$\psi^2 H^2 D$ & yes & direct & EOM & Figs.~\ref{fig:psi2H2D}, \ref{fig:X3} \\
\rowcolor{LightCyan}
$\psi^2 H D^2$ & no & --- & & \\
\midrule
\rowcolor{LightGray}
4-fermion operators & & & & \\
\midrule
$(\bar L L)(\bar L L)$ & yes & --- & EOM & Figure~\ref{fig:X3} \\
$(\bar R R)(\bar R R)$ & yes & --- & EOM & Figure~\ref{fig:X3} \\
$(\bar L L)(\bar R R)$ & yes & direct & EOM & Figs.~\ref{fig:psi4},\! \ref{fig:X3} \\
$(\bar L R)(\bar R L)$ & yes & direct & --- & Figure~\ref{fig:psi4} \\
$(\bar L R)(\bar L R)$ & yes & direct & --- & Figure~\ref{fig:psi4} \\
$B$-violating & yes & --- & --- & \\
\bottomrule
\end{tabular}
\caption{\label{tab:survey}
Summary of the different classes of dimension-6 operators in the extended SMEFT basis. $X$ represents a field-strength tensor (normal or dual), $H$ a Higgs field, $\psi$ a chiral fermion, and $D$ a covariant derivative. Operators contained in the Warsaw basis are shown on white background, while redundant operators are highlighted in blue. The third and fourth columns show which operators are generated by one-loop ALP exchange, either directly or via the EOMs. The last column refers to the figures showing the relevant Feynman diagrams. When more than one figure is listed, the first number refers to the figure showing the direct contributions.}
\end{table}

For a given Green's function, we express the sum over Feynman diagrams in the form
\begin{equation}
   \sum_i D_i^{\rm ALP} \equiv \frac{i\spac{\cal A}}{(4\pi f)^2} \,,
\end{equation}
where the coefficient of the UV $1/\epsilon$ pole, with $\epsilon=(4-d)/2$ being the dimensional regulator, is expressed in terms of matrix elements of operators in the extended basis. In order to remove these poles in the process of renormalization, the SMEFT operators are required as counterterms. This, in turn, implies that the ALP couplings appear as source terms in the RG evolution equations for the SMEFT Wilson coefficients, as shown in (\ref{SMEFTrge}). We prefer to perform our calculations using the second form of the effective ALP Lagrangian shown in (\ref{Leffalt}) rather than the original form (\ref{Leff}), which is completely equivalent. The reason is that in (\ref{Leffalt}) the ALP--fermion couplings always involve a Higgs field. It can then easily be seen without explicit calculation that there do not exist one-loop Feynman diagrams with ALP exchange that would require operators in the classes $X H^2 D^2$, $H^2 D^4$, $\psi^2 X D$, $\psi^2 D^3$ and $\psi^2 H D^2$ as counterterms. 

\subsection{Purely bosonic operators}

We begin by considering SMEFT operators built out of field-strength tensors $X$ (normal or dual), Higgs doublets $H$ and covariant derivatives $D$. The Warsaw basis contains operators in the classes $X^3$, $X^2 H^2$, $H^6$ and $H^4 D^2$, whereas all operators in the classes $X^2 D^2$, $X H^2 D^2$ and $H^2 D^4$ are redundant and can be reduced to the operators of the Warsaw basis using the EOMs. 

\paragraph{\boldmath Classes $X^3$ and $X^2 D^2$.}

The extended basis contains two types of operators in these classes: the Weinberg operators $Q_V$ and $Q_{\widetilde V}$ containing three gauge fields (with $V=G,W$), and operators containing two gauge fields and two covariant derivatives. The latter operators are redundant and can be eliminated using the EOMs, but they must be kept in the extended operator basis. Those involving gluons fields can be defined as 
$\widehat Q_{G,1}=\left( D_\rho\spac G_{\mu\nu} \right)^a \left( D^\rho\spac G^{\mu\nu} \right)^a$ and 
$\widehat Q_{G,2}=\left( D^\rho\spac G_{\rho\mu} \right)^a \left( D_\omega\spac G^{\omega\mu} \right)^a$. Here and below, operators $Q_i$ are contained in the Warsaw basis, while redundant operators in the extended basis are denoted by $\widehat Q_i$. The Bianchi identity implies the relation 
\begin{equation}
   2 g_s\spac Q_G + \widehat Q_{G,1} - 2\spac\widehat Q_{G,2} = 0 \,,
\end{equation}
which can be used to eliminate $\widehat Q_{G,1}$ from the extended basis. We are thus left with the operators
\begin{equation}\label{Ohatset1}
\begin{aligned}
   \widehat Q_{G,2} 
   &= \left( D^\rho\spac G_{\rho\mu} \right)^a \left( D_\omega\spac G^{\omega\mu} \right)^a , \\
   \widehat Q_{W,2} 
   &= \left( D^\rho\spac W_{\rho\mu} \right)^I \left( D_\omega\spac W^{\omega\mu} \right)^I , \\
   \widehat Q_{B,2} 
   &= \left( D^\rho\spac B_{\rho\mu} \right) \left( D_\omega\spac B^{\omega\mu} \right) .
\end{aligned}
\end{equation}
Analogous operators, in which one of the two field-strength tensors is replaced by a dual tensor, are not needed as counterterms in our analysis.

\begin{figure}
\centering
\includegraphics[width=0.714\textwidth]{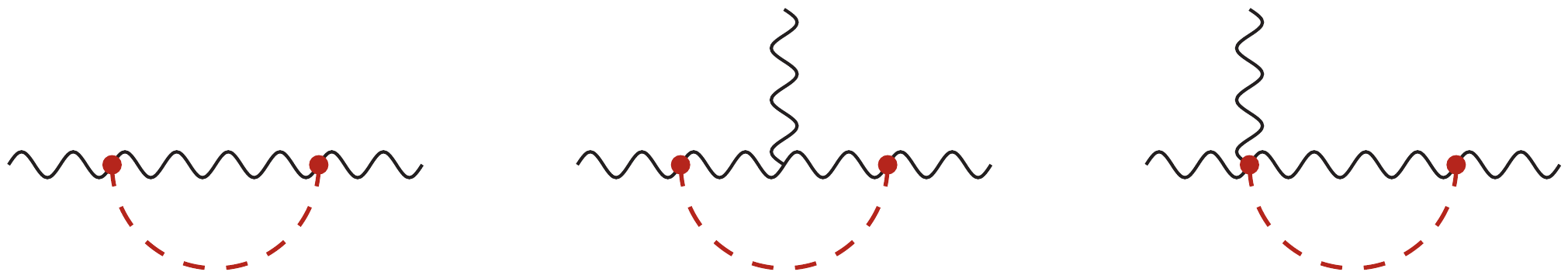}
\caption{\label{fig:X3} 
Representative one-loop Feynman diagrams with ALP exchange (red dashed line), which require operators in the classes $X^3$ and $X^2 D^2$ as counterterms. The 2-point functions (first graph) exist for all three types of gauge bosons, while the 3-point functions (last two graphs) require non-abelian vertices involving three gauge bosons.}
\end{figure}

At one-loop order, the 1PI Feynman diagrams with a virtual ALP exchange, which require operators in the classes $X^3$ and $X^2 D^2$ as counterterms, are shown in Figure~\ref{fig:X3}. Here and below, a red dashed line represents an ALP propagator, while red dots mark the $1/f$-suppressed ALP--SM vertices. In order to determine the coefficients of the counterterms we study both the three-boson and two-boson Green's functions with off-shell external momenta. The three-boson amplitudes only exist for the non-abelian gauge fields. Starting with the gluon case, we find that both the 3-gluon and the 2-gluon amplitude can be written in the form
\begin{equation}\label{3gres}
   {\cal A}\big(gg(g)\big) 
   = - \frac{C_{GG}^2}{\epsilon} \left[ 4\spac g_s\spac\langle Q_G\rangle 
    + \frac43\,\langle\widehat Q_{G,2}\rangle
    - 2 m_a^2\,\langle G_{\mu\nu}^a\spac G^{\mu\nu,a} \rangle \right] + \text{finite} \,,
\end{equation}
where the matrix element of $Q_G$ requires three external gluons to be non-zero. In a completely analogous way, we find that
\begin{equation}\label{3Wres}
\begin{aligned}
   {\cal A}\big(WW(W)\big) 
   &= - \frac{C_{WW}^2}{\epsilon} \left[ 4 g_2\spac\langle Q_W\rangle 
    + \frac43\,\langle\widehat Q_{W,2}\rangle
    - 2 m_a^2\,\langle W_{\mu\nu}^I\spac W^{\mu\nu,I} \rangle \right] + \text{finite} \,, \\
   {\cal A}(BB) 
   &= - \frac{C_{BB}^2}{\epsilon} \left[ \frac43\,\langle\widehat Q_{B,2}\rangle
    - 2 m_a^2\,\langle B_{\mu\nu}\spac B^{\mu\nu} \rangle \right] + \text{finite} \,.
\end{aligned}
\end{equation}
In all three cases, the presence of the contributions proportional to the ALP mass parameter $m_a^2$ leads to a wave-function renormalization of the gauge fields, which affects the scale evolution of the running couplings $\alpha_s(\mu)$, $\alpha_2(\mu)$ and $\alpha_1(\mu)$. This will be discussed in detail in Section~\ref{subsec:D=4ops}. 

\paragraph{\boldmath Classes $X^2 H^2$ and $X H^2 D^2$.}

\begin{figure}
\centering
\includegraphics[width=0.548\textwidth]{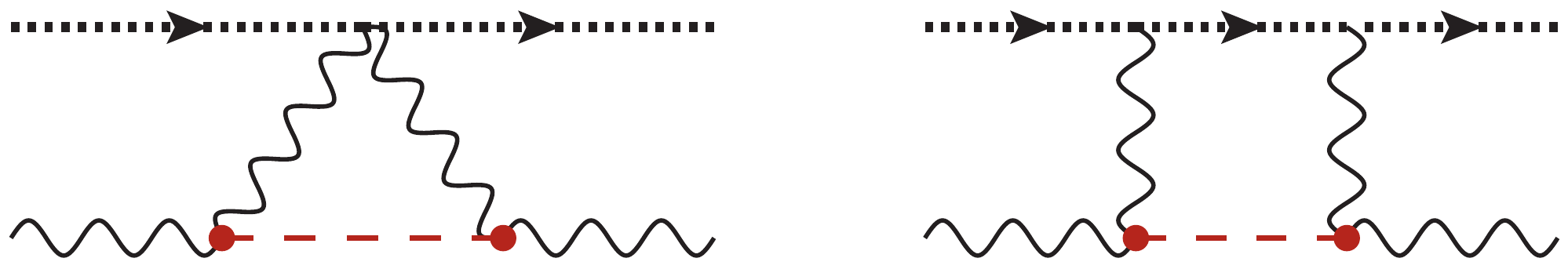}
\caption{\label{fig:X2H2} 
Representative one-loop Feynman diagrams with ALP exchange (red dashed line), which require operators in the class $X^2 H^2$ as counterterms. Higgs doublets are represented by thick dotted lines, with the arrow indicating the hypercharge flow.}
\end{figure}

At one-loop order, the 1PI Feynman diagrams with a virtual ALP exchange, which require operators in class $X^2 H^2$ as counterterms, are shown in Figure~\ref{fig:X2H2}. The vertices connecting Higgs and gauge bosons exist only for $SU(2)_L$ and $U(1)_Y$ gauge fields. We find that the UV divergences of the corresponding amplitudes (with all particles incoming) can be written in the form
\begin{equation}\label{VVHHres}
\begin{aligned}
   {\cal A}(WW H^*\negspac H) 
   &= \frac{C_{WW}^2}{\epsilon}\,g_2^2\,\langle Q_{HW}\rangle + \text{finite} \,, \\
   {\cal A}(BB H^*\negspac H) 
   &= \frac{C_{BB}^2}{\epsilon}\,g_1^2\,\langle Q_{HB}\rangle + \text{finite} \,, \\
   {\cal A}(W\negspac B H^*\negspac H) 
   &= \frac{C_{WW}\,C_{BB}}{\epsilon}\,2\spac g_1 g_2\,\langle Q_{HW\negspac B}\rangle 
    + \text{finite} \,.
\end{aligned}
\end{equation}
Operators in the class $X H^2 D^2$ are not generated by ALP exchange at one-loop order. 

\paragraph{\boldmath Classes $H^6$, $H^4 D^2$ and $H^2 D^4$.}

The operators in the classes $H^6$ and $H^4 D^2$ are not generated directly via one-loop diagrams involving ALP exchange, but they are generated indirectly when the redundant operators are eliminated using the EOMs. The relevant relations are derived in Section~\ref{subsec:redundantops}. Operators in the class $H^2 D^4$ are not generated by ALP exchange at one-loop order.

\subsection{Operators containing a single fermion current}

\paragraph{\boldmath Classes $\psi^2 X D$ and $\psi^2 D^3$.}

These operators are not generated by ALP exchange at one-loop order.

\paragraph{\boldmath Class $\psi^2 X H$.}

\begin{figure}
\centering
\includegraphics[width=0.548\textwidth]{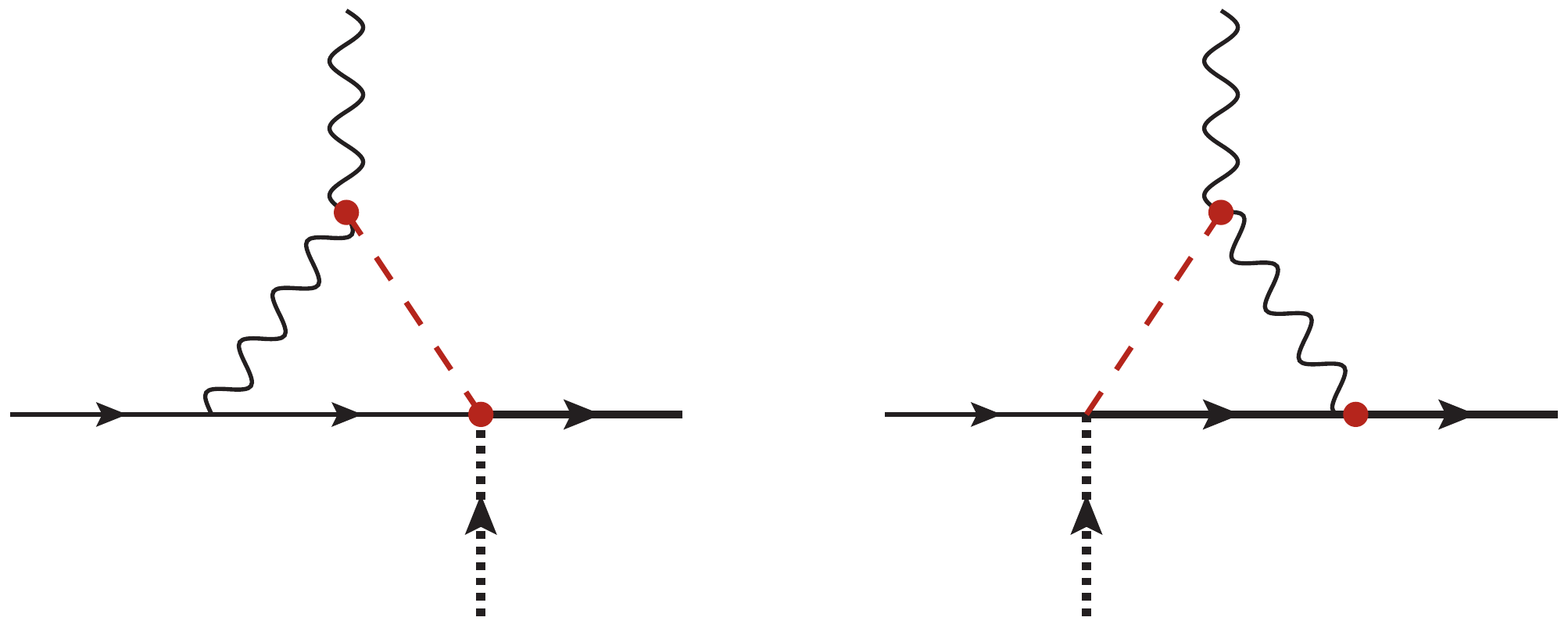}
\caption{\label{fig:psi2XH} 
Representative one-loop Feynman diagrams with ALP exchange (red dashed line), which require operators in the class $\psi^2 X H$ as counterterms. Thick (thin) solid lines represent left-handed fermion doublets (right-handed fermion singlets). If the right-handed fermion is an up-type quark, the arrows on the Higgs lines need to be reversed.}
\end{figure}

The operators in this class are generated by the one-loop Feynman graphs shown in Figure~\ref{fig:psi2XH}. We find that the UV divergences of these diagrams can be expressed as
\begin{align}\label{psi2XH}
   {\cal A}(L_p^*\spac e_r\spac W\negspac H) 
   &= \frac{i g_2}{2\epsilon}\,\big( \widetilde{\bm{Y}}_e \big)_{pr}\,
    C_{WW}\,\big[ \langle Q_{eW}\rangle \big]_{pr} + \text{finite} \,, \notag\\
   {\cal A}(L_p^*\spac e_r\spac B H) 
   &= \frac{i g_1}{\epsilon} \left( {\cal Y}_L + {\cal Y}_e \right) 
    \big( \widetilde{\bm{Y}}_e \big)_{pr}\,C_{BB}\,\big[ \langle Q_{eB}\rangle \big]_{pr} 
    + \text{finite} \,, \notag\\
   {\cal A}(Q_p^*\spac u_r\,g H^*) 
   &= \frac{2i g_s}{\epsilon}\,\big( \widetilde{\bm{Y}}_u \big)_{pr}\,
    C_{GG}\,\big[ \langle Q_{uG}\rangle \big]_{pr} + \text{finite} \,, \notag\\
   {\cal A}(Q_p^*\spac u_r\spac W\negspac H^*) 
   &= \frac{i g_2}{2\epsilon}\,\big( \widetilde{\bm{Y}}_u \big)_{pr}\,
    C_{WW}\,\big[ \langle Q_{uW}\rangle \big]_{pr} + \text{finite} \,, \notag\\
   {\cal A}(Q_p^*\spac u_r\spac B H^*) 
   &= \frac{i g_1}{\epsilon} \left( {\cal Y}_Q + {\cal Y}_u \right) 
    \big( \widetilde{\bm{Y}}_u \big)_{pr}\,C_{BB}\,\big[ \langle Q_{uB}\rangle \big]_{pr} 
    + \text{finite} \,, \\
   {\cal A}(Q_p^*\spac d_r\,g H) 
   &= \frac{2i g_s}{\epsilon}\,\big( \widetilde{\bm{Y}}_d \big)_{pr}\,
    C_{GG}\,\big[ \langle Q_{dG}\rangle \big]_{pr} + \text{finite} \,, \notag\\
   {\cal A}(Q_p^*\spac d_r\spac W H) 
   &= \frac{i g_2}{2\epsilon}\,\big( \widetilde{\bm{Y}}_d \big)_{pr}\,
    C_{WW}\,\big[ \langle Q_{dW}\rangle \big]_{pr} + \text{finite} \,, \notag\\
   {\cal A}(Q_p^*\spac d_r\spac B H) 
   &= \frac{i g_1}{\epsilon} \left( {\cal Y}_Q + {\cal Y}_d \right) 
    \big( \widetilde{\bm{Y}}_d \big)_{pr}\,C_{BB}\,\big[ \langle Q_{dB}\rangle \big]_{pr} 
    + \text{finite} \,, \notag
\end{align}
where $p,r=1,2,3$ are generation indices.

\paragraph{\boldmath Classes $\psi^2 H^3$, $\psi^2 H^2 D$ and $\psi^2 H D^2$.}

\begin{figure}[h]
\centering
\includegraphics[width=\textwidth]{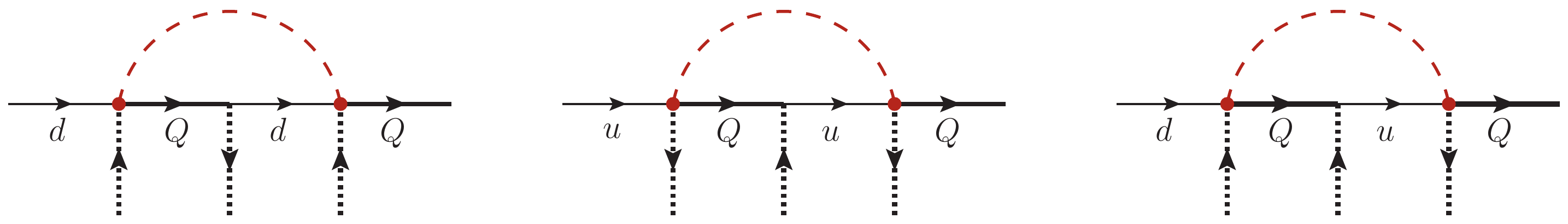}
\caption{\label{fig:psi2H3} 
Representative one-loop Feynman diagrams with ALP exchange (red dashed line), which require operators in the class $\psi^2 H^3$ as counterterms. Graphs involving both $\widetilde{\bm{Y}}_u$ and $\widetilde{\bm{Y}}_d$, such as the third one, vanish after summing over the permutations of the Higgs fields.}
\end{figure}

Operators in the class $\psi^2 H^3$ are generated by the one-loop Feynman graphs shown in Figure~\ref{fig:psi2H3}. We do not show a diagram analogous to the first one involving leptons. We find that the UV divergences of these diagrams can be expressed as
\begin{align}\label{psi2H3}
   {\cal A}(L_p^*\spac e_r\spac H^*\negspac HH) 
   &= \frac{1}{\epsilon}\,\big( \widetilde{\bm{Y}}_e\spac\bm{Y}_e^\dagger\spac
    \widetilde{\bm{Y}}_e \big)_{pr}\,\big[ \langle Q_{eH}\rangle \big]_{pr} + \text{finite} \,, \notag\\
   {\cal A}(Q_p^*\spac u_r\spac H^*\negspac H^*\negspac H) 
   &= \frac{1}{\epsilon}\,\big( \widetilde{\bm{Y}}_u\spac\bm{Y}_u^\dagger\spac
    \widetilde{\bm{Y}}_u \big)_{pr}\,\big[ \langle Q_{uH}\rangle \big]_{pr} + \text{finite} \,, \\
   {\cal A}(Q_p^*\spac d_r\spac H^*\negspac HH) 
   &= \frac{1}{\epsilon}\,\big( \widetilde{\bm{Y}}_d\spac\bm{Y}_d^\dagger\spac
    \widetilde{\bm{Y}}_d \big)_{pr}\,\big[ \langle Q_{dH}\rangle \big]_{pr} + \text{finite} \,. \notag
\end{align}
There exist diagrams such as the third one shown in the figure, which are proportional to structures like
$(\widetilde{\bm{Y}}_u\spac\bm{Y}_u^\dagger\spac\widetilde{\bm{Y}}_d \big)_{pr}$. However, we find that such graphs give vanishing contributions after summing over the permutations of the two incoming (or outgoing) Higgs bosons.

\begin{figure}
\centering
\includegraphics[width=0.714\textwidth]{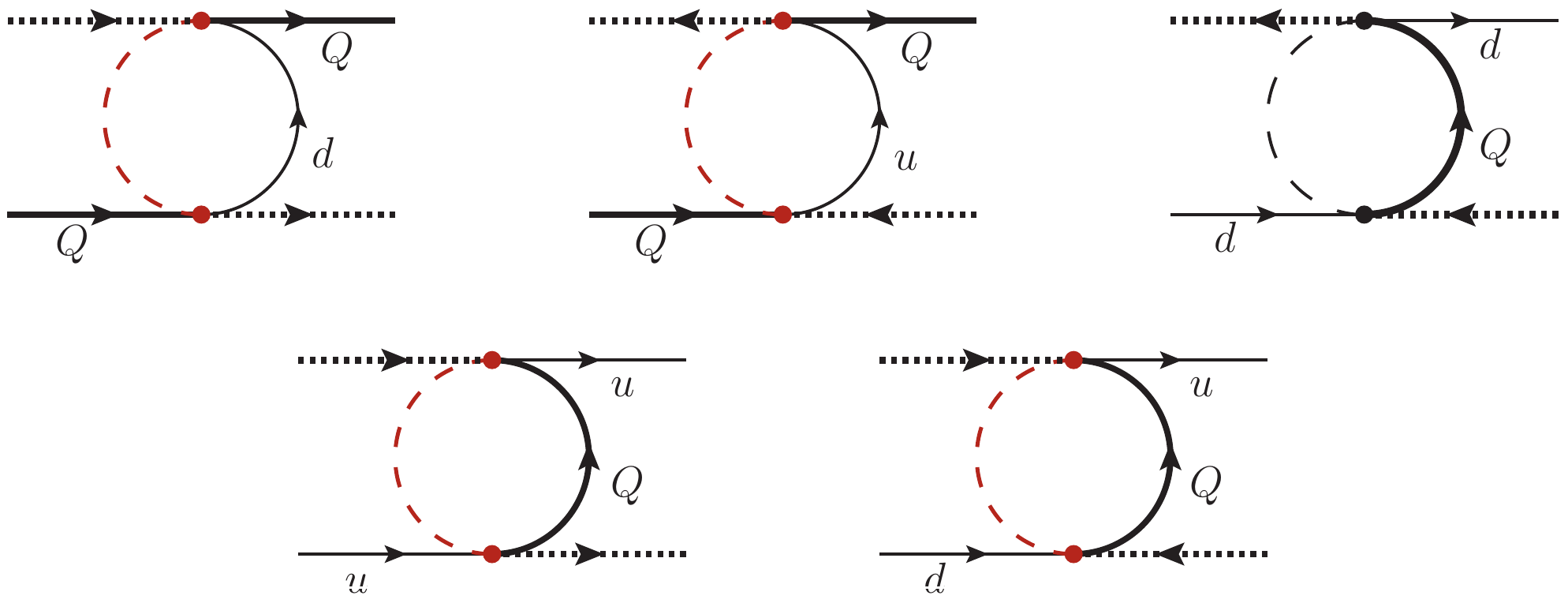}
\caption{\label{fig:psi2H2D} 
Representative one-loop Feynman diagrams with ALP exchange (red dashed line), which require operators in the class $\psi^2 H^2 D$ as counterterms. Analogous graphs exist in the lepton sector.}
\end{figure}

For the operators in the class $\psi^2 H^2 D$, which are generated by the one-loop Feynman graphs shown in Figure~\ref{fig:psi2H2D}, it is necessary to define the redundant operators 
\begin{equation}\label{Ohatset2}
\begin{aligned}
   \big[ \widehat Q_{Hl}^{(1)} \big]_{pr} 
   &= H^\dagger H\,\big( \bar L_p\,\iDslLR L_r \big) \,, &
   \big[ \widehat Q_{Hl}^{(3)} \big]_{pr} 
   &= H^\dagger\sigma^I H\,\big( \bar L_p\,\iDslLR\sigma^I L_r \big) \,, \\
   \big[ \widehat Q_{He} \big]_{pr} 
   &= H^\dagger H\,\big( \bar e_p\,\iDslLR e_r \big) \,, && \\
   \big[ \widehat Q_{Hq}^{(1)} \big]_{pr} 
   &= H^\dagger H\,\big( \bar Q_p\,\iDslLR Q_r \big) \,, &
   \big[ \widehat Q_{Hq}^{(3)} \big]_{pr} 
   &= H^\dagger\sigma^I H\,\big( \bar Q_p\,\iDslLR\sigma^I Q_r \big) \,, \\
   \big[ \widehat Q_{Hu} \big]_{pr} 
   &= H^\dagger H\,\big( \bar u_p\,\iDslLR u_r \big) \,, \qquad &
   \big[ \widehat Q_{Hd} \big]_{pr} 
   &= H^\dagger H\,\big( \bar d_p\,\iDslLR d_r \big) \,,
\end{aligned}
\end{equation}
which are not part of the Warsaw basis. They will later be eliminated using the EOMs. We find that the UV divergences of these diagrams can be expressed as 
\begin{align}\label{psi2H2D}
   {\cal A}(L_p^*\spac L_r\spac H^*\negspac H) 
   &= \frac{1}{8\epsilon}\,\big( \widetilde{\bm{Y}}_e\spac\widetilde{\bm{Y}}_e^\dagger \big)_{pr}
    \left( \big[ \langle \widehat Q_{Hl}^{(1)} \rangle \big]_{pr} 
    + \big[ \langle \widehat Q_{Hl}^{(3)} \rangle \big]_{pr} 
    - \big[ \langle Q_{Hl}^{(1)} \rangle \big]_{pr} 
    - \big[ \langle Q_{Hl}^{(3)} \rangle \big]_{pr} \right) + \text{finite} \,, \notag\\
   {\cal A}(e_p^*\spac e_r\spac H^*\negspac H) 
   &= \frac{1}{4\epsilon}\,\big( \widetilde{\bm{Y}}_e^\dagger\spac\widetilde{\bm{Y}}_e \big)_{pr}
    \left( \big[ \langle \widehat Q_{He} \rangle \big]_{pr} 
    + \big[ \langle Q_{He} \rangle \big]_{pr} \right) + \text{finite} \,, \notag\\
   {\cal A}(Q_p^*\spac Q_r\spac H^*\negspac H) 
   &= \frac{1}{8\epsilon}\,\big( \widetilde{\bm{Y}}_u\spac\widetilde{\bm{Y}}_u^\dagger \big)_{pr}
    \left( \big[ \langle \widehat Q_{Hq}^{(1)} \rangle \big]_{pr} 
    - \big[ \langle \widehat Q_{Hq}^{(3)} \rangle \big]_{pr} 
    + \big[ \langle Q_{Hq}^{(1)} \rangle \big]_{pr} 
    - \big[ \langle Q_{Hq}^{(3)} \rangle \big]_{pr} \right) + \text{finite} \,, \notag\\
   &+ \frac{1}{8\epsilon}\,\big( \widetilde{\bm{Y}}_d\spac\widetilde{\bm{Y}}_d^\dagger \big)_{pr}
    \left( \big[ \langle \widehat Q_{Hq}^{(1)} \rangle \big]_{pr} 
    + \big[ \langle \widehat Q_{Hq}^{(3)} \rangle \big]_{pr} 
    - \big[ \langle Q_{Hq}^{(1)} \rangle \big]_{pr} 
    - \big[ \langle Q_{Hq}^{(3)} \rangle \big]_{pr} \right) \\
   {\cal A}(u_p^*\spac u_r\spac H^*\negspac H) 
   &= \frac{1}{4\epsilon}\,\big( \widetilde{\bm{Y}}_u^\dagger\spac\widetilde{\bm{Y}}_u \big)_{pr}
    \left( \big[ \langle \widehat Q_{Hu} \rangle \big]_{pr} 
    - \big[ \langle Q_{Hu} \rangle \big]_{pr} \right) + \text{finite} \,, \notag\\
   {\cal A}(d_p^*\spac d_r\spac H^*\negspac H) 
   &= \frac{1}{4\epsilon}\,\big( \widetilde{\bm{Y}}_d^\dagger\spac\widetilde{\bm{Y}}_d \big)_{pr}
    \left( \big[ \langle \widehat Q_{Hd} \rangle \big]_{pr} 
    + \big[ \langle Q_{Hd} \rangle \big]_{pr} \right) + \text{finite} \,, \notag\\
   {\cal A}(u_p^*\spac d_r\spac H H) 
   &= \frac{1}{2\epsilon}\,\big( \widetilde{\bm{Y}}_u^\dagger\spac\widetilde{\bm{Y}}_d \big)_{pr}\,
    \big[ \langle Q_{Hud} \rangle \big]_{pr} + \text{finite} \,. \notag
\end{align}
Operators in class $\psi^2 H D^2$ are not generated by ALP exchange at one-loop order.

\subsection{Four-fermion operators}

\begin{figure}
\centering
\includegraphics[width=0.714\textwidth]{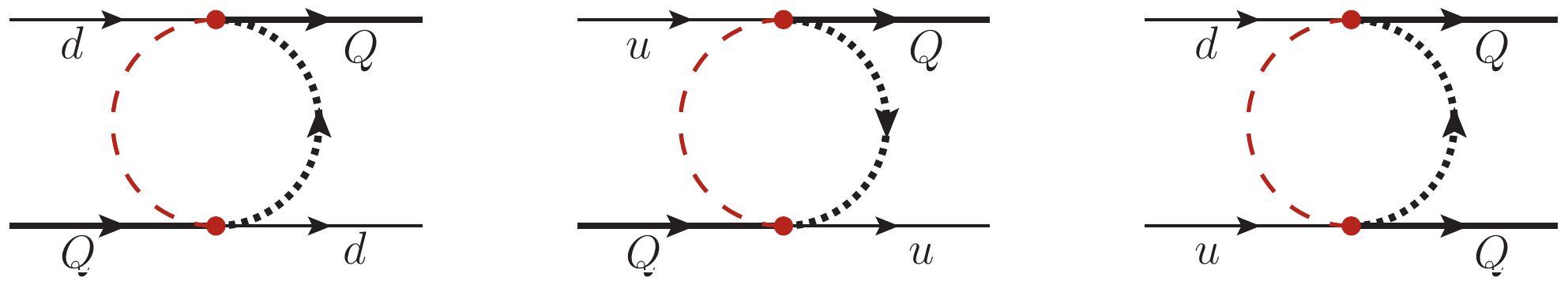}
\caption{\label{fig:psi4} 
Representative one-loop Feynman diagrams with ALP exchange (red dashed line), which require operators in the class $\psi^4$ as counterterms. Analogous graphs exist in the lepton sector.}
\end{figure}

At one-loop order, ALP exchange between four fermions gives rise to the diagrams shown in Figure~\ref{fig:psi4}. Since the ALP coupling to fermions changes chirality, each diagram contains two left-handed and two right-handed fermions. Four-fermion operators containing only left-handed or only right-handed fields are therefore not generated directly in our model at one-loop order. Nevertheless, as we will show later, almost all four-fermion operators in the Warsaw basis are generated at one-loop order when the contributions from the EOMs are taken into account. 

Using Fierz identities for some of the operators and color structures, we find that the amplitudes corresponding to the diagrams shown in Figure~\ref{fig:psi4} can be written as 
\begin{equation}\label{4fampls}
\begin{aligned}
   {\cal A}(L_p^*\spac e_r\spac e_s^*\spac L_t) 
   &= - \frac{1}{2\epsilon}\,\big( \widetilde{\bm{Y}}_e \big)_{pr} 
    \big( \widetilde{\bm{Y}}_e^\dagger \big)_{st}\,\big[ \langle Q_{le} \rangle \big]_{ptsr} 
    + \text{finite} \,, \\
   {\cal A}(Q_p^*\spac u_r\spac u_s^*\spac Q_t) 
   &= - \frac{1}{\epsilon}\,\big( \widetilde{\bm{Y}}_u \big)_{pr} 
    \big( \widetilde{\bm{Y}}_u^\dagger \big)_{st} \left(
    \big[ \langle Q_{qu}^{(8)} \rangle \big]_{ptsr} 
    + \frac{1}{2N_c}\,\big[ \langle Q_{qu}^{(1)} \rangle \big]_{ptsr} \right) + \text{finite} \,, \\
   {\cal A}(Q_p^*\spac d_r\spac d_s^*\spac Q_t) 
   &= - \frac{1}{\epsilon}\,\big( \widetilde{\bm{Y}}_d \big)_{pr} 
    \big( \widetilde{\bm{Y}}_d^\dagger \big)_{st} \left(
    \big[ \langle Q_{qd}^{(8)} \rangle \big]_{ptsr} 
    + \frac{1}{2N_c}\,\big[ \langle Q_{qd}^{(1)} \rangle \big]_{ptsr} \right) + \text{finite} \,, \\
   {\cal A}(L_p^*\spac e_r\spac d_s^*\spac Q_t) 
   &= \frac{1}{\epsilon}\,\big( \widetilde{\bm{Y}}_e \big)_{pr} 
    \big( \widetilde{\bm{Y}}_d^\dagger \big)_{st}\,\big[ \langle Q_{ledq} \rangle \big]_{prst} 
    + \text{finite} \,, \\
   {\cal A}(Q_p^*\spac d_r\spac Q_s^*\spac u_t) 
   &= \frac{1}{\epsilon}\,\big( \widetilde{\bm{Y}}_u \big)_{pr} 
    \big( \widetilde{\bm{Y}}_d \big)_{st}\,\big[ \langle Q_{quqd}^{(1)} \rangle \big]_{prst} 
    + \text{finite} \,, \\
   {\cal A}(L_p^*\spac e_r\spac Q_s^*\spac u_t) 
   &= - \frac{1}{\epsilon}\,\big( \widetilde{\bm{Y}}_e \big)_{pr} 
    \big( \widetilde{\bm{Y}}_u \big)_{st}\,\big[ \langle Q_{lequ}^{(1)} \rangle \big]_{prst} 
    + \text{finite} \,.
\end{aligned}
\end{equation}

\subsection{Elimination of redundant operators}
\label{subsec:redundantops}

In the next step, we must decompose the redundant operators $\widehat Q_i$ into SMEFT operators in the Warsaw basis, using the EOMs for the SM fields. We find that the relevant relations for the purely bosonic operators defined in (\ref{Ohatset1}) are 
\begin{equation}\label{QG2hat}
\begin{aligned}
   \widehat Q_{G,2} 
   &\cong g_s^2 \left( \bar Q\spac\gamma_\mu\spac T^a\spac Q + \bar u\spac\gamma_\mu\spac T^a\spac u
    + \bar d\spac\gamma_\mu\spac T^a\spac d \right)^2 \\
   &= g_s^2\,\bigg[\, \frac14 \left( \big[ Q_{qq}^{(1)} \big]_{prrp} + \big[ Q_{qq}^{(3)} \big]_{prrp} \right) 
    - \frac{1}{2 N_c}\,\big[ Q_{qq}^{(1)} \big]_{pprr}
    + \frac12\,\big[ Q_{uu} \big]_{prrp} - \frac{1}{2 N_c}\,\big[ Q_{uu} \big]_{pprr} \\
   &\hspace{12mm} + \frac12\,\big[ Q_{dd} \big]_{prrp} - \frac{1}{2 N_c}\,\big[ Q_{dd} \big]_{pprr}
    + 2\,\big[ Q_{qu}^{(8)} \big]_{pprr} + 2\,\big[ Q_{qd}^{(8)} \big]_{pprr}
    + 2\,\big[ Q_{ud}^{(8)} \big]_{pprr} \bigg] \,,
\end{aligned}
\end{equation}
\begin{equation}\label{QW2hat}
\begin{aligned}
   \widehat Q_{W,2} 
   &\cong \frac{g_2^2}{4} \left( H^\dagger\,i\hspace{-0.7mm}\overleftrightarrow{D}
    \hspace{-0.7mm}_\mu\hspace{-1mm}^I H + \bar Q\spac\gamma_\mu\spac\sigma^I Q 
    + \bar L\spac\gamma_\mu\spac\sigma^I L \right)^2 \\
   &= \frac{g_2^2}{4}\,\bigg[ - 4 m_H^2 \left( H^\dagger H \right)^2 + 4\lambda\,Q_H + 3\spac Q_{H\Box} 
    + 2 \left( \big[ Q_{Hl}^{(3)} \big]_{pp} + \big[ Q_{Hq}^{(3)} \big]_{pp} \right) \\
   &\hspace{13.1mm} + 2 \left[ \big( Y_u \big)_{pr}\,\big[ Q_{uH} \big]_{pr}
    + \big( Y_d \big)_{pr}\,\big[ Q_{dH} \big]_{pr} + \big( Y_e \big)_{pr}\,\big[ Q_{eH} \big]_{pr}
    + \text{h.c.} \right] \hspace{2.1cm}\\
   &\hspace{13.1mm} + 2\,\big[ Q_{lq}^{(3)} \big]_{pprr} 
    + 2\,\big[ Q_{ll} \big]_{prrp} - \big[ Q_{ll} \big]_{pprr} 
    + \big[ Q_{qq}^{(3)} \big]_{pprr} \bigg] \,,
\end{aligned}
\end{equation}
and
\begin{align}\label{QB2hat}
   \widehat Q_{B,2} 
   &\cong g_1^2\,\bigg( {\cal Y}_H\spac H^\dagger\,i\hspace{-0.7mm}\overleftrightarrow{D}
    \hspace{-0.7mm}_\mu H + \sum_F\,{\cal Y}_F\,\bar\psi_F\spac\gamma_\mu\spac\psi_F \bigg)^2 \notag\\
   &= g_1^2\,\bigg[\, {\cal Y}_H^2\,\big( 4\spac Q_{HD} + Q_{H\Box} \big) \notag\\[-1mm]
   &\hspace{12mm} + 2 {\cal Y}_H \left( {\cal Y}_L\,\big[ Q_{Hl}^{(1)} \big]_{pp}
    + {\cal Y}_Q\,\big[ Q_{Hq}^{(1)} \big]_{pp} + {\cal Y}_e\,\big[ Q_{He} \big]_{pp}
    + {\cal Y}_u\,\big[ Q_{Hu} \big]_{pp} + {\cal Y}_d\,\big[ Q_{Hd} \big]_{pp} \right) \hspace{3mm} \notag\\
   &\hspace{12mm} + {\cal Y}_L^2\,\big[ Q_{ll} \big]_{pprr} + {\cal Y}_Q^2\,\big[ Q_{qq}^{(1)} \big]_{pprr}
    + {\cal Y}_e^2\,\big[ Q_{ee} \big]_{pprr} + {\cal Y}_u^2\,\big[ Q_{uu} \big]_{pprr}
    + {\cal Y}_d^2\,\big[ Q_{dd} \big]_{pprr} \notag\\
   &\hspace{12mm} + 2 {\cal Y}_L\spac{\cal Y}_Q\,\big[ Q_{lq}^{(1)} \big]_{pprr}
    + 2 {\cal Y}_L\spac{\cal Y}_e\,\big[ Q_{le} \big]_{pprr}
    + 2 {\cal Y}_L\spac{\cal Y}_u\,\big[ Q_{lu} \big]_{pprr}
    + 2 {\cal Y}_L\spac{\cal Y}_d\,\big[ Q_{ld} \big]_{pprr} \notag\\
   &\hspace{12mm} + 2 {\cal Y}_Q\spac{\cal Y}_e\,\big[ Q_{qe} \big]_{pprr}
    + 2 {\cal Y}_Q\spac{\cal Y}_u\,\big[ Q_{qu}^{(1)} \big]_{pprr} 
    + 2 {\cal Y}_Q\spac{\cal Y}_d\,\big[ Q_{qd}^{(1)} \big]_{pprr} \notag\\[-1mm]
   &\hspace{12mm} + 2 {\cal Y}_e\spac{\cal Y}_u\,\big[ Q_{eu} \big]_{pprr} 
    + 2 {\cal Y}_e\spac{\cal Y}_d\,\big[ Q_{ed} \big]_{pprr}
    + 2 {\cal Y}_u\spac{\cal Y}_d\,\big[ Q_{ud}^{(1)} \big]_{pprr} \bigg] \,.
\end{align}
They are in agreement with corresponding relations derived in \cite{Gherardi:2020det}. In relation (\ref{QW2hat}), $m_H^2$ is the Higgs mass parameter and $\lambda$ the scalar self-coupling as defined via the scalar potential \cite{Grzadkowski:2010es}
\begin{equation}
   V = \frac{\lambda}{2} \left( H^\dagger H \right)^2 - m_H^2\,H^\dagger H \,.
\end{equation}
The relevant relations for the operators containing a single fermion current defined in (\ref{Ohatset2}) read
\begin{equation}\label{eq:25}
\begin{aligned}
   \big[ \widehat Q_{Hl}^{(1)} \big]_{pr} 
   &\cong \big( \bm{Y}_e \big)_{rs}\spac\big[ Q_{eH} \big]_{ps} 
    + \big( \bm{Y}_e^\dagger \big)_{sp}\spac\big[ Q_{eH}^\dagger \big]_{sr} \,, \\
   \big[ \widehat Q_{Hl}^{(3)} \big]_{pr} 
   &\cong \big( \bm{Y}_e \big)_{rs}\spac\big[ Q_{eH} \big]_{ps} 
    + \big( \bm{Y}_e^\dagger \big)_{sp}\spac\big[ Q_{eH}^\dagger \big]_{sr} \,, \\
   \big[ \widehat Q_{He} \big]_{pr} 
   &\cong \big( \bm{Y}_e \big)_{sp}\spac\big[ Q_{eH} \big]_{sr} 
    + \big( \bm{Y}_e^\dagger \big)_{rs}\spac\big[ Q_{eH}^\dagger \big]_{ps} \,, \\
   \big[ \widehat Q_{Hq}^{(1)} \big]_{pr} 
   &\cong \big( \bm{Y}_d \big)_{rs}\spac\big[ Q_{dH} \big]_{ps} 
    + \big( \bm{Y}_u \big)_{rs}\spac\big[ Q_{uH} \big]_{ps}
    + \big( \bm{Y}_d^\dagger \big)_{sp}\spac\big[ Q_{dH}^\dagger \big]_{sr} 
    + \big( \bm{Y}_u^\dagger \big)_{sp}\spac\big[ Q_{uH}^\dagger \big]_{sr} \,, \\
   \big[ \widehat Q_{Hq}^{(3)} \big]_{pr} 
   &\cong \big( \bm{Y}_d \big)_{rs}\spac\big[ Q_{dH} \big]_{ps} 
    - \big( \bm{Y}_u \big)_{rs}\spac\big[ Q_{uH} \big]_{ps}
    + \big( \bm{Y}_d^\dagger \big)_{sp}\spac\big[ Q_{dH}^\dagger \big]_{sr} 
    - \big( \bm{Y}_u^\dagger \big)_{sp}\spac\big[ Q_{uH}^\dagger \big]_{sr} \,, \\
   \big[ \widehat Q_{Hu} \big]_{pr} 
   &\cong \big( \bm{Y}_u \big)_{sp}\spac\big[ Q_{uH} \big]_{sr} 
    + \big( \bm{Y}_u^\dagger \big)_{rs}\spac\big[ Q_{uH}^\dagger \big]_{ps} \,, \\
   \big[ \widehat Q_{Hd} \big]_{pr} 
   &\cong \big( \bm{Y}_d \big)_{sp}\spac\big[ Q_{dH} \big]_{sr} 
    + \big( \bm{Y}_d^\dagger \big)_{rs}\spac\big[ Q_{dH}^\dagger \big]_{ps} \,. 
\end{aligned}
\end{equation}

\section{Derivation of the source terms}
\label{sec:4}

With the exception of the CP-violating bosonic operators as well as the baryon-number violating four-fermion operators, which do not arise in our model because the ALP neither has CP-violating couplings to gauge fields nor couplings that violate baryon number, almost all of the 59 SMEFT operators in the Warsaw basis are sourced by ALP exchange at one-loop order. There are only three special cases: the bosonic operator $Q_{HG}$ and the four-fermion operators $Q_{quqd}^{(8)}$ and $Q_{lequ}^{(3)}$ receive source terms starting at two-loop order. 

Below we list our results for the ALP source terms in the RG equations (\ref{SMEFTrge}) for the Wilson coefficients of the dimension-6 SMEFT Lagrangian in the various operator classes of the Warsaw basis. Before doing so, we consider ALP effects on the evolution equations for the SM coupling parameters, i.e., the couplings of the renormalizable ($D=4$) operators. 

\subsection{Renormalizable operators}
\label{subsec:D=4ops}

The effective Lagrangian (\ref{completeL}) contains two dimensionful parameters besides the scale $f$: the Higgs mass parameter $m_H^2$ and the ALP mass parameter $m_a^2$. One-loop diagrams with a virtual ALP exchange can therefore generate divergent terms proportional to $m_H^2/f^2$ or $m_a^2/f^2$, which require the dimension-4 operators of the SM as counterterms. We have seen two occurrences of this phenomenon: the contributions proportional to $m_a^2$ in (\ref{3gres}) and (\ref{3Wres}), and the term proportional to $m_H^2$ in (\ref{QW2hat}). A careful inspection of the UV-divergent Green's functions requiring SM counterterms shows that these are indeed the only such terms generated at one-loop order in the ALP model.

The divergent contributions proportional to $m_a^2$ in (\ref{3gres}) and (\ref{3Wres}) are absorbed by a wave-function renormalization of the gauge fields. The corresponding contributions to the $Z$ factors defined by $G_{\mu,0}^a=Z_G^{1/2}\,G_\mu^a$ etc.\ are
\begin{equation}
   \delta Z_G = \frac{8\spac m_a^2}{(4\pi f)^2}\,\frac{C_{GG}^2}{\epsilon} \,, \qquad
   \delta Z_W = \frac{8\spac m_a^2}{(4\pi f)^2}\,\frac{C_{WW}^2}{\epsilon} \,, \qquad
   \delta Z_B = \frac{8\spac m_a^2}{(4\pi f)^2}\,\frac{C_{BB}^2}{\epsilon} \,.
\end{equation}
These $Z$ factors enter in the relation between the bare and the renormalized gauge couplings, and consequently the presence of the ALP leaves an imprint on the $\beta$-functions of the SM gauge couplings. For the case of the running QCD coupling, $\alpha_s\equiv\alpha_s(\mu)$, the relation is $\alpha_{s,0}=\mu^{2\epsilon}\spac Z_{\alpha_s}\,\alpha_s$ with $Z_{\alpha_s}=Z_{\bar qq g}^2\,Z_q^{-2}\,Z_G^{-1}$. The wave-function renormalization factor $Z_q$ of the quark fields and the vertex renormalization factor $Z_{\bar qq g}$ do not receive ALP contributions at one-loop order. The fact that the bare coupling $\alpha_{s,0}$ is scale independent implies the relation 
\begin{equation}
   \frac{1}{\alpha_s}\,\frac{d\spac\alpha_s}{d\ln\mu}
   = - 2\epsilon - \frac{1}{Z_{\alpha_s}}\,\frac{d\spac Z_{\alpha_s}}{d\ln\mu} \,, \quad
    \text{with} \quad
   Z_{\alpha_s} = Z_{\alpha_s}^{\rm SM} - \frac{8\spac m_a^2}{(4\pi f)^2}\,\frac{C_{GG}^2}{\epsilon} \,.
\end{equation}
In evaluating the derivative of $Z_{\alpha_s}$ one uses that $d\spac\alpha_s/d\ln\mu=-2\epsilon\spac\alpha_s+\dots$ and $d\spac C_{GG}^2/d\ln\mu=-2\epsilon\spac C_{GG}^2+\dots$ up to higher-order corrections. The latter relation follows from the fact that the bare coupling $C_{GG,0}$ in the effective Lagrangian (\ref{Leffalt}) has mass dimension $[C_{GG,0}]=\epsilon$. Taking the limit $\epsilon\to 0$ after the derivatives have been computed, and defining $d\spac\alpha_s/d\ln\mu\equiv-2\alpha_s\spac\beta^{(3)}(\{\alpha_i\})$, and similarly for the other gauge couplings, we obtain
\begin{equation}
\begin{aligned}
   \beta^{(3)}(\{\alpha_i\}) 
   &= \beta_{\rm SM}^{(3)}(\{\alpha_i\}) + \frac{8\spac m_a^2}{(4\pi f)^2}\,C_{GG}^2 \,, \\
   \beta^{(2)}(\{\alpha_i\}) 
   &= \beta_{\rm SM}^{(2)}(\{\alpha_i\}) + \frac{8\spac m_a^2}{(4\pi f)^2}\,C_{WW}^2 \,, \\
   \beta^{(1)}(\{\alpha_i\}) 
   &= \beta_{\rm SM}^{(1)}(\{\alpha_i\}) + \frac{8\spac m_a^2}{(4\pi f)^2}\,C_{BB}^2 \,.
\end{aligned}
\end{equation}
The notation $\{\alpha_i\}$ indicates a dependence on all coupling parameters in the SM, which enters starting at two-loop order.

The divergent contribution proportional to $m_H^2$ in (\ref{QW2hat}) is absorbed by the renormalization of the quartic scalar coupling $\lambda$. The renormalization factor $Z_\lambda$, defined by $\lambda_0=\mu^{2\epsilon}\spac Z_\lambda\,\lambda(\mu)$, receives a contribution
\begin{equation}
   \delta Z_\lambda 
   = \frac{8\spac g_2^2}{3\spac\lambda}\,\frac{m_H^2}{(4\pi f)^2}\,\frac{C_{WW}^2}{\epsilon} \,.
\end{equation}
This generates an additive, ALP-induced contribution to the RG evolution equation for the renormalized coupling $\lambda(\mu)$, such that
\begin{equation}
   \frac{d\spac\lambda}{d\ln\mu}
   = - \frac{16\spac g_2^2}{3}\,\frac{m_H^2}{(4\pi f)^2}\,C_{WW}^2 
    + \text{SM contribution} \,.
\end{equation}

\subsection{Source terms of purely bosonic operators}

We now turn our attention to the derivation of the ALP source terms in the RG evolution equation (\ref{SMEFTrge}). The dimension-5 operators in the effective ALP Lagrangian ${\cal L}_{\rm SM+ALP}$ in (\ref{completeL}) give rise to UV-divergent Green's functions, which require the dimension-6 SMEFT operators $Q_i$ as counterterms. More specifically, the bare Wilson coefficients $C_{i,0}$ in the dimension-6 SMEFT Lagrangian
\begin{equation}
   {\cal L}_{\rm SMEFT}^{D=6} = \sum_i\,C_{i,0}\,Q_{i,0}
\end{equation}
must contain divergent contributions, which cancel the $1/\epsilon$ poles of the ALP contributions. Consider the contribution proportional to the Weinberg operator $Q_G$ in the 3-gluon amplitude (\ref{3gres}) as an example. In order to cancel the corresponding $1/\epsilon$ pole, the bare Wilson coefficient $C_{G,0}$ must contain the contribution
\begin{equation}
   C_{G,0} \ni \frac{4\spac g_s}{(4\pi f)^2}\,C_{GG}^2
    \left( \frac{1}{\epsilon} + \ln\frac{\mu^2}{M^2} + \dots \right) ,
\end{equation}
where $M^2$ is a characteristic mass scale in the UV theory, and the combination of $1/\epsilon$ and $\ln\mu^2$ is generic for one-loop integrals in dimensional regularization. When the Wilson coefficient is renormalized, the $1/\epsilon$ pole term is removed, but the scale-dependent term remains. It follows that
\begin{equation}
   \frac{d}{d\ln\mu}\,C_G(\mu) \ni \frac{8\spac g_s}{(4\pi f)^2}\,C_{GG}^2 \,.
\end{equation}
In this way, the ALP source terms for the various Wilson coefficients can be derived from the coefficients of the $1/\epsilon$ poles in the expressions for the various divergent Green's functions considered in Section~\ref{sec:3}.

\vspace{3mm}
\noindent
\underline{Class $X^3$:}\\[2mm]
\noindent
From the results for the $1/\epsilon$ poles in the two- and three-point gauge-boson amplitudes shown in (\ref{3gres}) and (\ref{3Wres}), we obtain the ALP source terms
\begin{equation}\label{Weinbergsources}
\begin{aligned}
   S_G &= 8\spac g_s\spac C_{GG}^2 \,, &
   S_{\widetilde G} &= 0 \,, \\
   S_W &= 8\spac g_2\spac C_{WW}^2 \,, \,\,\quad & 
   S_{\widetilde W} &= 0 \,.
\end{aligned}
\end{equation}

\vspace{3mm}
\noindent
\underline{Class $X^2 H^2$:}\\[2mm]
\noindent
From the results for the $1/\epsilon$ poles in the amplitudes connecting two Higgs bosons and two gauge fields shown in (\ref{VVHHres}), we obtain the ALP source terms
\begin{equation}
\begin{aligned}
   S_{HG} &= 0 \,, &
   S_{H\widetilde G} &= 0 \,, \\
   S_{HW} &= - 2\spac g_2^2\,C_{WW}^2 \,, &
   S_{H\widetilde W} &= 0 \,, \\
   S_{HB} &= - 2\spac g_1^2\,C_{BB}^2 \,, &
   S_{H\widetilde B} &= 0 \,, \\
   S_{HW\negspac B} &= - 4\spac g_1 g_2\,C_{WW}\spac C_{BB} \,, \,\,\quad &
   S_{H\widetilde WB} &= 0 \,.
\end{aligned}
\end{equation}
Here and in (\ref{Weinbergsources}), the source terms for the CP-odd operators (marked with a tilde) vanish at one-loop order, because the ALP does not have any CP-violating couplings to gauge bosons.

\vspace{3mm}
\noindent
\underline{Classes $H^6$ and $H^4 D^2$:}\\[2mm]
\noindent
The operators in these classes do not receive any direct contributions from one-loop diagrams with ALP exchange, but they are generated via contributions from the EOMs due to the operators $\widehat Q_{W,2}$ and $\widehat Q_{B,2}$, see relations (\ref{QW2hat}) and (\ref{QB2hat}). We find
\begin{equation}
\begin{aligned}
   S_H &= \blueterms{ \frac83\,\lambda\spac g_2^2\,C_{WW}^2 } \,, \\
   S_{H\Box} &= \blueterms{ 2\spac g_2^2\,C_{WW}^2 }
    + \blueterms{ \frac83\,g_1^2\,{\cal Y}_H^2\,C_{BB}^2 } \,, \\
   S_{HD} &= \blueterms{ \frac{32}{3}\,g_1^2\,{\cal Y}_H^2\,C_{BB}^2 } \,.
\end{aligned}
\end{equation}

\subsection{Source terms of single fermion-current operators}

The Wilson coefficients of these operators are matrices in generation space. We present our results for the corresponding source term using a matrix notation with boldface symbols.

\vspace{3mm}
\noindent
\underline{Class $\psi^2 X H$:}\\[2mm]
From the results for the $1/\epsilon$ poles in the amplitudes connecting two fermions to a Higgs field and a gauge field shown in (\ref{psi2XH}), we obtain the ALP source terms
\begin{equation}\label{dipolesources}
\begin{aligned}
   \bm{S}_{eW} &= - i g_2\,\widetilde{\bm{Y}}_e\,C_{WW} \,, \\
   \bm{S}_{eB} &= - 2i g_1 \left( {\cal Y}_L + {\cal Y}_e \right) \widetilde{\bm{Y}}_e\,C_{BB} \,, \\
   \bm{S}_{uG} &= - 4i g_s\,\widetilde{\bm{Y}}_u\,C_{GG} \,, \\
   \bm{S}_{uW} &= - i g_2\,\widetilde{\bm{Y}}_u\,C_{WW} \,, \\
   \bm{S}_{uB} &= - 2i g_1 \left( {\cal Y}_Q + {\cal Y}_u \right) \widetilde{\bm{Y}}_u\,C_{BB} \,, \\
   \bm{S}_{dG} &= - 4i g_s\,\widetilde{\bm{Y}}_d\,C_{GG} \,, \\
   \bm{S}_{dW} &= - i g_2\,\widetilde{\bm{Y}}_d\,C_{WW} \,, \\
   \bm{S}_{dB} &= - 2i g_1 \left( {\cal Y}_Q + {\cal Y}_d \right) \widetilde{\bm{Y}}_d\,C_{BB} \,.
\end{aligned}
\end{equation}

\vspace{3mm}
\noindent
\underline{Class $\psi^2 H^3$:}\\[2mm]
The source terms for the operators in this class receive direct contributions, as shown in (\ref{psi2H3}), as well as contributions from EOMs, from the relations given in (\ref{QW2hat}) and (\ref{eq:25}). We find
\begin{equation}
\begin{aligned}
   \bm{S}_{eH} &= - 2\spac\widetilde{\bm{Y}}_e\spac\bm{Y}_e^\dagger\spac\widetilde{\bm{Y}}_e 
    - \blueterms{ \frac12\spac\widetilde{\bm{Y}}_e\spac\widetilde{\bm{Y}}_e^\dagger\spac\bm{Y}_e } 
    - \blueterms{ \frac12\spac\bm{Y}_e\spac\widetilde{\bm{Y}}_e^\dagger\spac\widetilde{\bm{Y}}_e } 
    + \blueterms{ \frac43\,g_2^2\,C_{WW}^2\spac\bm{Y}_e } \,, \\
   \bm{S}_{uH} &= - 2\spac\widetilde{\bm{Y}}_u\spac\bm{Y}_u^\dagger\spac\widetilde{\bm{Y}}_u 
    - \blueterms{ \frac12\spac\widetilde{\bm{Y}}_u\spac\widetilde{\bm{Y}}_u^\dagger\spac\bm{Y}_u } 
    - \blueterms{ \frac12\spac\bm{Y}_u\spac\widetilde{\bm{Y}}_u^\dagger\spac\widetilde{\bm{Y}}_u } 
    + \blueterms{ \frac43\,g_2^2\,C_{WW}^2\spac\bm{Y}_u } \,, \\
   \bm{S}_{dH} &= - 2\spac\widetilde{\bm{Y}}_d\spac\bm{Y}_d^\dagger\spac\widetilde{\bm{Y}}_d 
    - \blueterms{ \frac12\spac\widetilde{\bm{Y}}_d\spac\widetilde{\bm{Y}}_d^\dagger\spac\bm{Y}_d } 
    - \blueterms{ \frac12\spac\bm{Y}_d\spac\widetilde{\bm{Y}}_d^\dagger\spac\widetilde{\bm{Y}}_d } 
    + \blueterms{ \frac43\,g_2^2\,C_{WW}^2\spac\bm{Y}_d } \,.
\end{aligned}
\end{equation}

\vspace{11mm}
\noindent
\underline{Class $\psi^2 H^2 D$:}\\[2mm]
The source terms for the operators in this class receive direct contributions, as shown in (\ref{psi2H2D}), as well as contributions from EOMs, from the relations given in (\ref{QW2hat}) and (\ref{QB2hat}). We find
\begin{equation}\label{eq:36}
\begin{aligned}
   \bm{S}_{Hl}^{(1)} &= \frac14\,\widetilde{\bm{Y}}_e\spac\widetilde{\bm{Y}}_e^\dagger 
    + \blueterms{ \frac{16}{3}\,g_1^2\,{\cal Y}_H\spac{\cal Y}_L\,C_{BB}^2\,\bm{1} } \,, \\
   \bm{S}_{Hl}^{(3)} &= \frac14\,\widetilde{\bm{Y}}_e\spac\widetilde{\bm{Y}}_e^\dagger 
    + \blueterms{ \frac43\,g_2^2\,C_{WW}^2\,\bm{1} } \,, \\
   \bm{S}_{He} &= - \frac12\,\widetilde{\bm{Y}}_e^\dagger\spac\widetilde{\bm{Y}}_e 
    + \blueterms{ \frac{16}{3}\,g_1^2\,{\cal Y}_H\spac{\cal Y}_e\,C_{BB}^2\,\bm{1} } \,, \\
   \bm{S}_{Hq}^{(1)} &= \frac14 \left( \widetilde{\bm{Y}}_d\spac\widetilde{\bm{Y}}_d^\dagger 
    - \widetilde{\bm{Y}}_u\spac\widetilde{\bm{Y}}_u^\dagger \right) 
    + \blueterms{ \frac{16}{3}\,g_1^2\,{\cal Y}_H\spac{\cal Y}_Q\,C_{BB}^2\,\bm{1} } \,, \\
   \bm{S}_{Hq}^{(3)} &= \frac14 \left( \widetilde{\bm{Y}}_d\spac\widetilde{\bm{Y}}_d^\dagger 
    + \widetilde{\bm{Y}}_u\spac\widetilde{\bm{Y}}_u^\dagger \right) 
    + \blueterms{ \frac43\,g_2^2\,C_{WW}^2\,\bm{1} } \,, \\
   \bm{S}_{Hu} &= \frac12\,\widetilde{\bm{Y}}_u^\dagger\spac\widetilde{\bm{Y}}_u 
    + \blueterms{ \frac{16}{3}\,g_1^2\,{\cal Y}_H\spac{\cal Y}_u\,C_{BB}^2\,\bm{1} } \,, \\
   \bm{S}_{Hd} &= - \frac12\,\widetilde{\bm{Y}}_d^\dagger\spac\widetilde{\bm{Y}}_d 
    + \blueterms{ \frac{16}{3}\,g_1^2\,{\cal Y}_H\spac{\cal Y}_d\,C_{BB}^2\,\bm{1} } \,, \\
   \bm{S}_{Hud} &= - \widetilde{\bm{Y}}_u^\dagger\spac\widetilde{\bm{Y}}_d \,.
\end{aligned}
\end{equation}

\subsection{Source terms of four-fermion operators}

The Wilson coefficients of these operators are 4-index tensors in generation space, and we therefore present our results for the corresponding source term in component notation. The direct contributions to the source terms are derived form the four-fermion amplitudes collected in (\ref{4fampls}). In addition, there are several indirect contributions from the EOM relations in (\ref{QG2hat}), (\ref{QW2hat}) and (\ref{QB2hat}). The source terms for operators in the classes $(\bar LL)(\bar LL)$ and $(\bar RR)(\bar RR)$ are entirely due to these EOM relations.

\vspace{3mm}
\noindent
\underline{Class $(\bar LL)(\bar LL)$:}\\[2mm]
For the source terms of the purely left-handed four-fermion operators we obtain
\begin{equation}\label{eq:37}
\begin{aligned}
   \big[ S_{ll} \big]_{prst} 
   &= \blueterms{ \frac23\,g_2^2\,C_{WW}^2  
    \left( 2\spac\delta_{pt}\spac\delta_{sr} - \delta_{pr}\spac\delta_{st} \right) } 
    + \blueterms{ \frac83\,g_1^2\,{\cal Y}_L^2\,C_{BB}^2\,\delta_{pr}\spac\delta_{st} } \,, \\
   \big[ S_{qq}^{(1)} \big]_{prst} 
   &= \blueterms{ \frac23\,g_s^2\,C_{GG}^2 
    \left( \delta_{pt}\spac\delta_{sr} - \frac{2}{N_c}\,\delta_{pr}\spac\delta_{st} \right) } 
    + \blueterms{ \frac83\,g_1^2\,{\cal Y}_Q^2\,C_{BB}^2\,\delta_{pr}\spac\delta_{st} } \,, \\
   \big[ S_{qq}^{(3)} \big]_{prst} 
   &= \blueterms{ \frac23\,g_s^2\,C_{GG}^2\,\delta_{pt}\spac\delta_{sr} } 
    + \blueterms{ \frac23\,g_2^2\,C_{WW}^2\,\delta_{pr}\spac\delta_{st} } \,, \\
   \big[ S_{lq}^{(1)} \big]_{prst} 
   &= \blueterms{ \frac{16}{3}\,g_1^2\,{\cal Y}_L\spac{\cal Y}_Q\,C_{BB}^2\, 
    \delta_{pr}\spac\delta_{st} } \,, \\
   \big[ S_{lq}^{(3)} \big]_{prst} 
   &= \blueterms{ \frac43\,g_2^2\,C_{WW}^2\,\delta_{pr}\spac\delta_{st} } \,.
\end{aligned}
\end{equation}

\vspace{3mm}
\noindent
\underline{Class $(\bar RR)(\bar RR)$:}\\[2mm]
For the source terms of the purely right-handed four-fermion operators we obtain
\begin{equation}
\begin{aligned}
   \big[ S_{ee} \big]_{prst} 
   &= \blueterms{ \frac83\,g_1^2\,{\cal Y}_e^2\,C_{BB}^2\,\delta_{pr}\spac\delta_{st} } \,, \\
   \big[ S_{uu} \big]_{prst} 
   &= \blueterms{ \frac43\,g_s^2\,C_{GG}^2 
    \left( \delta_{pt}\spac\delta_{sr} - \frac{1}{N_c}\,\delta_{pr}\spac\delta_{st} \right) } 
    + \blueterms{ \frac83\,g_1^2\,{\cal Y}_u^2\,C_{BB}^2\,\delta_{pr}\spac\delta_{st} } \,, \\
   \big[ S_{dd} \big]_{prst} 
   &= \blueterms{ \frac43\,g_s^2\,C_{GG}^2 
    \left( \delta_{pt}\spac\delta_{sr} - \frac{1}{N_c}\,\delta_{pr}\spac\delta_{st} \right) } 
    + \blueterms{ \frac83\,g_1^2\,{\cal Y}_d^2\,C_{BB}^2\,\delta_{pr}\spac\delta_{st} } \,, \\
   \big[ S_{eu} \big]_{prst} 
   &= \blueterms{ \frac{16}{3}\,g_1^2\,{\cal Y}_e\spac{\cal Y}_u\,C_{BB}^2\,
    \delta_{pr}\spac\delta_{st} } \,, \\
   \big[ S_{ed} \big]_{prst} 
   &= \blueterms{ \frac{16}{3}\,g_1^2\,{\cal Y}_e\spac{\cal Y}_d\,C_{BB}^2\,
    \delta_{pr}\spac\delta_{st} } \,, \\
   \big[ S_{ud}^{(1)} \big]_{prst} 
   &= \blueterms{ \frac{16}{3}\,g_1^2\,{\cal Y}_u\spac{\cal Y}_d\,C_{BB}^2\,
    \delta_{pr}\spac\delta_{st} } \,, \\
   \big[ S_{ud}^{(8)} \big]_{prst} 
   &= \blueterms{ \frac{16}{3}\,g_s^2\,C_{GG}^2\,\delta_{pr}\spac\delta_{st} } \,.
\end{aligned}
\end{equation}

\vspace{3mm}
\noindent
\underline{Class $(\bar LL)(\bar RR)$:}\\[2mm]
For the source terms of the mixed-chirality four-fermion operators in this class we obtain
\begin{equation}
\begin{aligned}
   \big[ S_{le} \big]_{prst} 
   &= \big( \widetilde{\bm{Y}}_e \big)_{pt} \big( \widetilde{\bm{Y}}_e^\dagger \big)_{sr} 
    + \blueterms{ \frac{16}{3}\,g_1^2\,{\cal Y}_L\spac{\cal Y}_e\,C_{BB}^2\, 
    \delta_{pr}\spac\delta_{st} } \,, \\
   \big[ S_{lu} \big]_{prst} 
   &= \blueterms{ \frac{16}{3}\,g_1^2\,{\cal Y}_L\spac{\cal Y}_u\,C_{BB}^2\, 
    \delta_{pr}\spac\delta_{st} } \,, \\
   \big[ S_{ld} \big]_{prst} 
   &= \blueterms{ \frac{16}{3}\,g_1^2\,{\cal Y}_L\spac{\cal Y}_d\,C_{BB}^2\, 
    \delta_{pr}\spac\delta_{st} } \,, \\
   \big[ S_{qe} \big]_{prst} 
   &= \blueterms{ \frac{16}{3}\,g_1^2\,{\cal Y}_Q\spac{\cal Y}_e\,C_{BB}^2\, 
    \delta_{pr}\spac\delta_{st} } \,, \\
   \big[ S_{qu}^{(1)} \big]_{prst} 
   &= \frac{1}{N_c}\,\big( \widetilde{\bm{Y}}_u \big)_{pt} \big( \widetilde{\bm{Y}}_u^\dagger \big)_{sr} 
    + \blueterms{ \frac{16}{3}\,g_1^2\,{\cal Y}_Q\spac{\cal Y}_u\,C_{BB}^2\, 
    \delta_{pr}\spac\delta_{st} } \,, \\
   \big[ S_{qu}^{(8)} \big]_{prst} 
   &= 2\spac\big( \widetilde{\bm{Y}}_u \big)_{pt} \big( \widetilde{\bm{Y}}_u^\dagger \big)_{sr} 
    + \blueterms{ \frac{16}{3}\,g_s^2\,C_{GG}^2\,\delta_{pr}\spac\delta_{st} } \,, \\
   \big[ S_{qd}^{(1)} \big]_{prst} 
   &= \frac{1}{N_c}\,\big( \widetilde{\bm{Y}}_d \big)_{pt} \big( \widetilde{\bm{Y}}_d^\dagger \big)_{sr} 
    + \blueterms{ \frac{16}{3}\,g_1^2\,{\cal Y}_Q\spac{\cal Y}_d\,C_{BB}^2\,
    \delta_{pr}\spac\delta_{st} } \,, \\
   \big[ S_{qd}^{(8)} \big]_{prst} 
   &= 2\spac\big( \widetilde{\bm{Y}}_d \big)_{pt} \big( \widetilde{\bm{Y}}_d^\dagger \big)_{sr} 
    + \blueterms{ \frac{16}{3}\,g_s^2\,C_{GG}^2\,\delta_{pr}\spac\delta_{st} } \,.
\end{aligned}
\end{equation}

\vspace{13mm}
\noindent
\underline{Classes $(\bar LR)(\bar RL)$ and $(\bar LR)(\bar LR)$:}\\[2mm]
For the source terms of the mixed-chirality four-fermion operators in these classes we obtain
\begin{equation}
\begin{aligned}
   \big[ S_{ledq} \big]_{prst} 
   &= - 2\spac\big( \widetilde{\bm{Y}}_e \big)_{pr} \big( \widetilde{\bm{Y}}_d^\dagger \big)_{st} \,, \\
   \big[ S_{quqd}^{(1)} \big]_{prst} 
   &= - 2\spac\big( \widetilde{\bm{Y}}_u \big)_{pr} \big( \widetilde{\bm{Y}}_d \big)_{st} \,, \\
   \big[ S_{quqd}^{(8)} \big]_{prst} 
   &= 0 \,, \\
   \big[ S_{lequ}^{(1)} \big]_{prst} 
   &= 2\spac\big( \widetilde{\bm{Y}}_e \big)_{pr} \big( \widetilde{\bm{Y}}_u \big)_{st} \,, \\
   \big[ S_{lequ}^{(3)} \big]_{prst} 
   &= 0 \,.
\end{aligned}
\end{equation}

\vspace{3mm}
\noindent
\underline{Class $B$-violating:}\\[2mm]
The $B$-violating operators $Q_{duq}$, $Q_{qqu}$, $Q_{qqq}$ and $Q_{duu}$ are not generated in the ALP model, because the model does not contain any $B$-violating interactions.

\subsection{Structure of the source terms}
\label{sec:4.5}

It is instructive to study the structure of the various ALP source terms in more detail. For the bosonic ALP couplings $C_{VV}$ with $V=G,W,B$, we have presented in (\ref{eq:6}) the relations which link them with the couplings in the underlying shift-symmetric ALP Lagrangian (\ref{Leff}). Note that, besides the three original ALP--boson couplings $c_{VV}$, also the diagonal elements of all ALP--fermion couplings enter in these relations. In other words, even in so-called gauge-phobic models, in which some or all of the original ALP--boson couplings are assumed to vanish, the couplings $C_{VV}$ in the ALP source terms are nevertheless non-zero as soon as the ALP has at least some couplings to the SM fermions.

The fermionic ALP couplings in the source terms are encoded in the complex matrices $\widetilde{\bm{Y}}_f$ with $f=u,d,e$ defined in (\ref{rela1}). They inherit the hierarchies of the SM Yukawa matrices $\bm{Y}_f$, which multiply the hermitian matrices $\bm{c}_F$ in the original Lagrangian (\ref{Leff}). We can simplify the structure of the matrices $\widetilde{\bm{Y}}_f$ by choosing a convenient basis of the fermion fields. Without loss of generality, we work in the basis where the up-sector and lepton-sector Yukawa matrices are diagonal, while the down-sector Yukawa matrix is given by $\bm{Y}_d=\bm{V}\,\bm{Y}_d^{\rm diag}$ with $\bm{Y}_d^{\rm diag}=\text{diag}(y_d,y_s,y_b)$, where $\bm{V}$ denotes the CKM matrix. Following \cite{Bauer:2017ris,Bauer:2020jbp}, we denote the ALP--fermion couplings in this basis by $\bm{k}_U=\bm{c}_Q$, $\bm{k}_E=\bm{c}_L$, and $\bm{k}_f=\bm{c}_f$ for $f=u,d,e$. Moreover, we define $\bm{k}_D=\bm{V}^\dagger\bm{c}_Q\bm{V}$. With these definitions, the matrices $\bm{k}_i$ specify the ALP--fermion couplings in the mass basis of the SM fermions. From (\ref{rela1}), it the follows that
\begin{equation}\label{eq:44}
\begin{aligned}
   -i\spac\big[\widetilde{\bm{Y}}_u]_{ij} 
   &= y_{u_i} \big[\bm{k}_u\big]_{ij} - \big[\bm{k}_U\big]_{ij}\,y_{u_j} \,, \\
   -i\spac\big[\widetilde{\bm{Y}}_d\big]_{ij} 
   &= \left( V_{u_i d_k}\,y_{d_k} \big[\bm{k}_d\big]_{kj} 
     - \big[ \bm{k}_U \big]_{ik}\spac V_{u_k d_j}\,y_{d_j} \right)
    = V_{u_i d_k}\!\left( y_{d_k} \big[\bm{k}_d\big]_{kj} - \big[\bm{k}_D\big]_{kj}\,y_{d_j} \right) , \\
   -i \spac\big[\widetilde{\bm{Y}}_e\big]_{ij} 
   &= y_{e_i} \big[\bm{k}_e\big]_{ij} - \big[\bm{k}_E\big]_{ij}\,y_{e_j} \,.
\end{aligned}
\end{equation}
The diagonal elements of these matrices are given by
\begin{equation}\label{eq:45}
\begin{aligned}
   -i\spac\big[\widetilde{\bm{Y}}_u]_{ii} 
   &= c_{u_i u_i}\,y_{u_i} \,, \\
   -i\spac\big[\widetilde{\bm{Y}}_d\big]_{ii} 
   &= V_{u_i d_k}\!\left( y_{d_k} \big[\bm{k}_d\big]_{ki} - \big[\bm{k}_D\big]_{ki}\,y_{d_i} \right) , \\
   -i \spac\big[\widetilde{\bm{Y}}_e\big]_{ii} 
   &= c_{e_i e_i}\,y_{e_i} \,,
\end{aligned}
\end{equation}
where we have defined \cite{Bauer:2017ris}
\begin{equation}
   c_{f_i f_i} = \big[ \bm{k}_f \big]_{ii} - \big[ \bm{k}_F \big]_{ii} \,.
\end{equation}

Further simplifications arise if one makes assumptions about the flavor structure of the ALP--fermion couplings in (\ref{Leff}). For example, assuming minimal flavor violation (MFV) \cite{DAmbrosio:2002vsn}, and neglecting contributions quadratic in the Yukawa couplings of the light SM fermions ($f\ne t$), one finds that the matrices $\bm{c}_u$ and $\bm{c}_Q$ are diagonal but in general have non-universal 33 entries, whereas $\bm{c}_d$, $\bm{c}_e$ and $\bm{c}_L$ are proportional to the unit matrix \cite{Bauer:2020jbp}. The unitarity of the CKM matrix then implies that 
\begin{equation}
\begin{aligned}
   \big[ \bm{k}_D \big]_{11}
   &= \big[ \bm{k}_U \big]_{11}
    + \left| V_{td} \right|^2 \left( \big[ \bm{k}_U \big]_{33} - \big[ \bm{k}_U \big]_{11} \right) , \\
   \big[ \bm{k}_D \big]_{22}
   &= \big[ \bm{k}_U \big]_{22}
    + \left| V_{ts} \right|^2 \left( \big[ \bm{k}_U \big]_{33} - \big[ \bm{k}_U \big]_{11} \right) , \\
   \big[ \bm{k}_D \big]_{33}
   &= \big[ \bm{k}_U \big]_{33}
    - \left( 1 - \left| V_{tb} \right|^2 \right) 
    \left( \big[ \bm{k}_U \big]_{33} - \big[ \bm{k}_U \big]_{11} \right) .
\end{aligned}
\end{equation}
In the Wolfenstein parameterization of the CKM matrix one has $\left|V_{td}\right|^2\sim\lambda^6$, $\left|V_{ts}\right|^2\sim\lambda^4$ and $\left( 1 - \left| V_{tb} \right|^2 \right)\sim\lambda^4$, where $\lambda\sim 0.2$. It follows that we can replace $\big[\bm{k}_U\big]_{ii}=\big[\bm{k}_D\big]_{ii}$ to very good approximation. Under the MFV hypothesis, the relations (\ref{eq:44}) then simplify to 
\begin{equation}
\begin{aligned}
   -i\spac\big[\widetilde{\bm{Y}}_u]_{ij}^{\rm MFV} 
   &\simeq c_{u_i u_i}\,y_{u_i}\,\delta_{ij} \,, \\
   -i\spac\big[\widetilde{\bm{Y}}_d\big]_{ij}^{\rm MFV} 
   &\simeq c_{d_i d_i}\spac V_{u_i d_j}\,y_{d_j} \,, \\
   -i \spac\big[\widetilde{\bm{Y}}_e\big]_{ij}^{\rm MFV}
   &\simeq c_{e_i e_i}\,y_{e_i}\,\delta_{ij} \,.
\end{aligned}
\end{equation}
In matrix notation, these relations imply $-i\spac\widetilde{\bm{Y}}_f^{\rm MFV}\simeq\text{diag}(c_{f_1 f_1},c_{f_2 f_2},c_{f_3 f_3})\,\bm{Y}_f$ for all three cases ($f=u,d,e$).

\section{Sample applications}
\label{sec:applications}

The results obtained in this paper form the basis of a systematic analysis of the effects of virtual ALP exchange on precision measurements. They also explain the origin of the logarithmic divergences observed in some previous studies of ALP-induced loop corrections to the anomalous magnetic moment of the muon \cite{Marciano:2016yhf,Bauer:2017nlg,Bauer:2017ris,Bauer:2019gfk,Buen-Abad:2021fwq} and electroweak precision observables \cite{Bauer:2017ris}, and they provide a framework in which such logarithms can be resummed by solving the RG evolution equations (\ref{SMEFTrge}). Many of the dimension-6 operators which receive contributions from ALP-induced RG evolution effects are strongly constrained, for example by measurements of electroweak precision observables and of the properties of the Higgs boson, the top quark and the gauge bosons at the LHC (see \cite{Ellis:2020unq} for a comprehensive global analysis and an exhaustive list of references to earlier SMEFT fits). This implies that areas of the ALP parameter space which may still be unconstrained by direct searches can be probed indirectly, using constraints on dimension-6 SMEFT operators implied by precision studies. We now briefly illustrate the usefulness of our approach with two examples, leaving a more comprehensive analysis to future work. For the purposes of this discussion we assume that the ALP mass is light, of order the electroweak scale or lighter. In this first exploration we neglect the matching contributions to the SMEFT Wilson coefficients from heavy new states at the UV scale $\Lambda=4\pi f$, which can only be assessed within a concrete UV completion of the effective Lagrangian (\ref{completeL}). We also omit one-loop contributions to the observables arising from the low-energy matrix elements in the effective theory. As explained earlier, the effects from RG evolution which we calculate are enhanced over these two contributions by a large logarithm. Our calculations have shown that the same ALP couplings appear in the source terms for many different dimension-6 operators, so it is likely that more powerful constraints than the ones we discuss below can be derived from a global analysis of precision observables.

\subsection{Chromo-magnetic moment of the top quark}

The chromo-magnetic and chromo-electric dipole moments of the top quark, $\hat\mu_t$ and $\hat d_t$, are two important precision observables probing new physics above the electroweak scale \cite{Atwood:1994vm,Haberl:1995ek,Cheung:1995nt}. They can be defined in terms of the effective Lagrangian \cite{Haberl:1995ek}
\begin{equation}
   {\cal L}_{t\bar t g}
   = g_s\,\bigg( \bar t\spac\gamma^\mu\spac T^a\spac t\,G_\mu^a
    + \frac{\hat\mu_t}{2m_t}\,\bar t\,\sigma^{\mu\nu}\spac T^a\spac t\,G_{\mu\nu}^a
    + \frac{i\spac\hat d_t}{2m_t}\,\bar t\,\sigma^{\mu\nu} \gamma_5\,T^a\spac t\,G_{\mu\nu}^a \bigg) \,.
\end{equation}
The overall sign on the right-hand side has been chosen so as to be consistent with our definition of the covariant derivative. Matching this expression with the dimension-6 SMEFT Lagrangian at lowest order, we find
\begin{equation}
   \hat\mu_t = \frac{y_t\spac v^2}{g_s}\,{\Re}e\spac C_{uG}^{33} \,, \qquad
   \hat d_t = \frac{y_t\spac v^2}{g_s}\,{\Im}m\spac C_{uG}^{33} \,, 
\end{equation}
where all quantities are evaluated at the scale $\mu=m_t$. The Wilson coefficient $C_{uG}^{33}\equiv\left[C_{uG}\right]_{33}$ is defined in the up-quark mass basis (see Section~\ref{sec:4.5}). Neglecting contributions proportional to electroweak gauge couplings and light-quark Yukawa couplings, one finds that this coefficient obeys the RG equation \cite{Jenkins:2013wua,Alonso:2013hga}
\begin{equation}
\begin{aligned}
   \frac{d}{d\ln\mu}\,C_{uG}^{33}
   &= \frac{S_{uG}^{33}}{(4\pi f)^2} 
    + \left( \frac{15\spac\alpha_t}{8\pi} - \frac{17\alpha_s}{12\pi} \right) C_{uG}^{33} \\
   &\quad + \frac{9\spac\alpha_s}{4\pi}\,y_t\!\left( C_G + i\spac C_{\widetilde G} \right) 
    + \frac{g_s\spac y_t}{4\pi^2} \left( C_{HG} + i\spac C_{H\widetilde G} \right) ,
\end{aligned}
\end{equation}
where $S_{uG}^{33}\equiv\left[\bm{S}_{uG}\right]_{33}$, and $\alpha_t=y_t^2/(4\pi)$. In the same approximation, the RG equations for the other Wilson coefficients entering this relation read
\begin{equation}
\begin{aligned}
   \frac{d}{d\ln\mu}\,C_G
   &= \frac{S_G}{(4\pi f)^2} + \frac{15\spac\alpha_s}{4\pi}\,C_G \,, \\
   \frac{d}{d\ln\mu}\,C_{\widetilde G}
   &= \frac{15\spac\alpha_s}{4\pi}\,C_{\widetilde G} \,, \\
   \frac{d}{d\ln\mu}\,C_{HG}
   &= \left( \frac{3\spac\alpha_t}{2\pi} - \frac{7\alpha_s}{2\pi} \right) C_{HG} 
    + \frac{g_s\spac y_t}{4\pi^2}\,\spac{\Re}e\spac C_{uG}^{33} \,, \\
   \frac{d}{d\ln\mu}\,C_{H\widetilde G}
   &= \left( \frac{3\spac\alpha_t}{2\pi} - \frac{7\alpha_s}{2\pi} \right) C_{H\widetilde G} 
    + \frac{g_s\spac y_t}{4\pi^2}\,\spac{\Im}m\spac C_{uG}^{33} \,.
\end{aligned}
\end{equation}
The relevant ALP source terms, 
\begin{equation}
   S_{uG}^{33} = 4\spac g_s\spac y_t\,c_{tt}\,C_{GG} \,, \qquad
   S_G = 8\spac g_s\,C_{GG}^2 \,,
\end{equation}
obtained from (\ref{Weinbergsources}), (\ref{dipolesources}) and (\ref{eq:45}), are both real-valued. It follows that $C_{\widetilde G}$, $C_{H\widetilde G}$ and ${\Im}m\spac C_{uG}^{33}$ vanish in the ALP model, and the RG equations simply to 
\begin{equation}
\begin{aligned}
   \frac{d}{d\ln\mu}\,{\Re}e\spac C_{uG}^{33}
   &= \frac{S_{uG}^{33}}{(4\pi f)^2} 
    + \left( \frac{15\spac\alpha_t}{8\pi} - \frac{17\alpha_s}{12\pi} \right) {\Re}e\spac C_{uG}^{33} 
    + \frac{9\spac\alpha_s}{4\pi}\,y_t\,C_G + \frac{g_s\spac y_t}{4\pi^2}\,C_{HG} \,, \\
   \frac{d}{d\ln\mu}\,C_G
   &= \frac{S_G}{(4\pi f)^2} + \frac{15\spac\alpha_s}{4\pi}\,C_G \,, \\
   \frac{d}{d\ln\mu}\,C_{HG}
   &= \left( \frac{3\spac\alpha_t}{2\pi} - \frac{7\alpha_s}{2\pi} \right) C_{HG} 
    + \frac{g_s\spac y_t}{4\pi^2}\,\spac{\Re}e\spac C_{uG}^{33} \,.
\end{aligned}
\end{equation}
Solving these coupled equations would provide solutions for the Wilson coefficients in which the large logarithms of the ratio $4\pi f/m_t$ are resummed in leading logarithmic approximation. For our purposes, however, it will be sufficient to obtain a rough approximation by keeping only the lowest-order logarithmic term for each ALP coupling and neglecting contributions proportional to extra factors of $\alpha_i\ln(4\pi f/m_t)$. Taking into account the RG equation for the coefficient $c_{tt}$ that follows from (\ref{eq:B.3}), we find
\begin{equation}
\begin{aligned}
   \hat\mu_t
   &\approx - \frac{8\spac m_t^2}{(4\pi f)^2} \left[ c_{tt}\,C_{GG}\,\ln\frac{4\pi f}{m_t}
    - \frac{25\spac\alpha_s}{4\pi}\,C_{GG}^2\,\ln^2\frac{4\pi f}{m_t} \right] \\
   &\approx - \left( 5.87\spac c_{tt}\,C_{GG} - 5.50\,C_{GG}^2 \right)\cdot 10^{-3}
    \times \left[ \frac{1\,\text{TeV}}{f} \right]^2 , 
\end{aligned}
\end{equation}
as well as $\hat d_t\approx 0$. The ALP couplings $c_{tt}$ and $C_{GG}$ are defined at the scale $\Lambda=4\pi f$. Note that the term proportional to $C_{GG}^2$ contains an extra factor of $\alpha_s\ln(4\pi f/m_t)$ compared with the first one, since it arises via the mixing of $C_{uG}^{33}$ with the coefficient $C_G$. The numerical result shown in the second line has been obtained using $m_t\equiv m_t(m_t)=163.4$\,GeV and $\alpha_s\equiv\alpha_s(m_t)=0.1084$, and taking $f=1$\,TeV in the argument of the logarithms. Which one of the two contributions dominates depends on the relative size of the coefficients $c_{tt}$ and $C_{GG}$. The CMS collaboration has recently performed two independent measurements of the chromo-magnetic dipole moment of the top quark, finding $-0.014<\hat\mu_t<0.004$ at 95\% confidence level \cite{CMS:2018jcg}, and $\hat\mu_t=-0.024\spac_{-0.009}^{+0.013}\spac_{-0.011}^{+0.016}$ \cite{Sirunyan:2019eyu}. Applying the (stronger) first bound to the ALP model, we find under the approximations described above
\begin{equation}
   - 0.68 
   < \left( c_{tt}\,C_{GG} - 0.94\,C_{GG}^2 \right)\times\left[ \frac{1\,\text{TeV}}{f} \right]^2 
   < 2.38 \qquad \text{(95\% CL)} \,.
\end{equation}
With the current sensitivity, the measurements of the top-quark chromo-magnetic moment probe the ALP couplings $c_{tt}/f$ and $C_{GG}/f$ at the level of roughly ${\cal O}(\text{TeV}^{-1})$.

\subsection[Example of a $Z$-pole constraint]{\boldmath Example of a $Z$-pole constraint}

As a second example, we consider the constraint on the flavor-conserving part of the Wilson coefficient of the dimension-6 SMEFT operator $Q_{Hq}^{(3)}$. Focussing on light quark flavors, and assuming flavor universality in the first two generations, one defines
\begin{equation}
   C_{Hq}^{(3)}\equiv \big[C_{Hq}^{(3)}\big]_{11} = \big[C_{Hq}^{(3)}\big]_{22} \,.
\end{equation}
The coefficient $C_{Hq}^{(3)}$ is tightly constrained by $Z$-pole measurements \cite{Han:2004az,Pomarol:2013zra,Ellis:2014jta,Falkowski:2014tna,Berthier:2015gja}. When marginalizing over all the other SMEFT coefficients in order to obtain the most conservative bound, the global analysis presented in \cite{Ellis:2020unq} yields
\begin{equation}\label{CMSbound}
   - 0.11\,\text{TeV}^{-2} < C_{Hq}^{(3)} < 0.012\,\spac\text{TeV}^{-2} \qquad \text{(95\% CL)} \,.
\end{equation}
Neglecting again contributions proportional to electroweak gauge couplings and light-quark Yukawa couplings, the RG equations for the coefficients $\big[C_{Hq}^{(3)}\big]_{ii}$ with $i\ne 3$ are found to be \cite{Jenkins:2013wua,Alonso:2013hga} 
\begin{equation}\label{RGECHq3}
\begin{aligned}
   \frac{d}{d\ln\mu}\,\big[C_{Hq}^{(3)}\big]_{ii}
   &= \frac{\big[S_{Hq}^{(3)}\big]_{ii}}{(4\pi f)^2} 
    + \frac{3\spac\alpha_t}{2\pi}\,\big[C_{Hq}^{(3)}\big]_{ii} 
    - \frac{3\spac\alpha_t}{2\pi}\,
    \Big( \big[C_{qq}^{(3)}\big]_{ii33} + \big[C_{qq}^{(3)}\big]_{33ii} \Big) \\
   &\quad - \frac{\alpha_t}{4\pi}\,\Big( 
    \big[C_{qq}^{(1)}\big]_{i33i} + \big[C_{qq}^{(1)}\big]_{3ii3} 
    - \big[C_{qq}^{(3)}\big]_{i33i} - \big[C_{qq}^{(3)}\big]_{3ii3} \Big) \,.
\end{aligned}
\end{equation}
From (\ref{eq:36}), we find for the relevant ALP source terms
\begin{equation}\label{SHq3}
   \big[ S_{Hq}^{(3)} \big]_{ii} 
   = \frac{4\spac g_2^2}{3}\,C_{WW}^2 \,,
\end{equation}
where again we have set the light-quark Yukawa couplings to zero. The source terms for the remaining operators in (\ref{RGECHq3}) are obtained from (\ref{eq:37}) and read (for $i=1,2$)
\begin{equation}
\begin{aligned}
   \big[S_{qq}^{(3)}\big]_{ii33} = \big[S_{qq}^{(3)}\big]_{33ii} 
   &= \frac{2\spac g_2^2}{3}\,C_{WW}^2 \,, \\
   \big[S_{qq}^{(1)}\big]_{i33i} = \big[S_{qq}^{(1)}\big]_{3ii3}
   &= \frac{2\spac g_s^2}{3}\,C_{GG}^2 \,, \\
   \big[S_{qq}^{(3)}\big]_{i33i} = \big[S_{qq}^{(3)}\big]_{3ii3}
   &= \frac{2\spac g_s^2}{3}\,C_{GG}^2 \,.
\end{aligned}
\end{equation}
They are indeed flavor universal with respect to the first two generations. We observe that the sum of all source terms (proportional to $C_{GG}^2$) for the Wilson coefficients in the second line of (\ref{RGECHq3}) vanishes. The source terms for the last two coefficients in the first line are proportional to $C_{WW}^2$ and thus yield a higher-order logarithmic correction to the contributions from the source term in (\ref{SHq3}), which we neglect. We thus find
\begin{equation}
   C_{Hq}^{(3)}\approx - \frac{\alpha}{3\pi\sin^2\theta_w}\,\frac{C_{WW}^2}{f^2}\,\ln\frac{4\pi f}{m_Z}
   \approx - 0.0177\,\frac{C_{WW}^2}{f^2} \,, 
\end{equation}
where the numerical value in the last step has been obtained using $\alpha(m_Z)=(127.95)^{-1}$, $\sin^2\theta_w=0.2312$ and setting $f=1$\,TeV in the argument of the logarithms. The lower bound in (\ref{CMSbound}) implies the limit 
\begin{equation}\label{CWWbound}
   |C_{WW}|\times \left[ \frac{1\,\text{TeV}}{f} \right] < 2.50 \qquad \text{(95\% CL)} \,.
\end{equation}
For ALPs in the mass range between 10\,MeV and 10\,GeV, and assuming that the ALP--photon coupling arises only from the ALP coupling to $SU(2)_L$ gauge bosons (i.e.\ that $C_{BB}=0$), one finds that this bound is of the same order of magnitude as the strongest constraints on the ALP--photon coupling obtained from LEP-2 \cite{Abbiendi:2002je,Knapen:2016moh} and Belle-II data \cite{BelleII:2020fag} (see also \cite{Bauer:2017ris}). With a future high-luminosity lepton collider such as the FCC-ee, the bound (\ref{CWWbound}) could be improved significantly  \cite{Bauer:2018uxu}.

Clearly, using the marginalized bound (\ref{CMSbound}) on a single operator as a proxy for a full fit is a very crude approach, especially given that we expect effects from $C_{WW}$ in several operators which enter into the same observables used in the fit. It is likely that a full analysis will lead to a significantly stronger bound on $C_{WW}$ and allow for more robust statements also in the case where several ALP couplings are non-zero. Performing such a global analysis is beyond the scope of the current work but is an interesting direction for future study.

\section{Conclusions}
\label{sec:conclusions}

While the Standard Model Effective Field Theory (SMEFT) is commonly used to describe the effects of new particles, which are too heavy to be produced as on-shell resonances in current high-energy physics experiments, we have shown that light new particles that are weakly coupled to the SM via non-renormalizable interactions necessarily induce non-zero Wilson coefficients in the SMEFT Lagrangian via renormalization-group evolution. The reason is that loop diagrams involving a virtual exchange of such a light particle contain UV divergences, which require higher-dimensional SMEFT operators (operators built out of SM fields only) as counterterms. 

For the well-motivated example of axions and axion-like particles (ALPs) interacting with the SM via dimension-5 interactions, we have computed the one-loop divergences of all Green's functions with virtual ALP exchange. We have expressed the results in terms of the matrix elements of dimension-6 SMEFT operators, carefully accounting for the contributions of redundant operators, which are eliminated using the equations of motion. In this way, we have derived the complete set of ALP source terms in the one-loop renormalization-group equations of the Wilson coefficients of the dimension-6 SMEFT operators in the Warsaw basis, finding that almost all of these operators are sourced at one-loop order in ALP interactions. Our results explain the origin of the UV divergences observed in previous one-loop calculations of virtual ALP contributions to the anomalous magnetic moment of the muon and to electroweak precision observables. More generally, they capture in a model-independent way all possible UV divergences of ALP-induced transitions between SM particles that can arise at one-loop order. Moreover, our formalism provides a convenient tool for resumming the large logarithmic corrections remaining once these UV divergences have been renormalized. 

As two important applications of our method, we have studied the ALP contributions to the chromo-magnetic dipole moment of the top quark and to electroweak precision tests performed at the $Z$ pole, finding that these observables provide interesting constraints on the ALP couplings to gauge bosons and top quarks if the ALP decay constant lies in the TeV region, and hence the scale $\Lambda=4\pi f$ is of order 10\,TeV. A more comprehensive analysis of virtual ALP effects based on a global fit to precision data is left for future work.
More generally, our formalism offers a model-independent framework for studying virtual ALP effects on a large variety of precision measurements, thus opening up new, indirect ways to search for ALPs and place bounds on their couplings to the SM. This approach is complementary to direct searches and independent of the way in which the ALP decays. 

\subsubsection*{Acknowledgments}

M.N.~thanks Gino Isidori, the particle theory group at Zurich University and the Pauli Center for hospitality during a sabbatical stay. The research of A.M.G.\ and M.N.\ was supported by the Cluster of Excellence {\em Precision Physics, Fundamental Interactions and Structure of Matter\/} (PRISMA${}^+$ -- EXC~2118/1) within the German Excellence Strategy (project ID 39083149). S.R.\ acknowledges support from the INFN grant no.~SESAMO.

\begin{appendix}

\section{\boldmath ALP source terms in an alternative operator basis}
\label{app:A}
\renewcommand{\theequation}{A.\arabic{equation}}
\setcounter{equation}{0}

In order to avoid odd powers of the gauge couplings $g_i$ and mixed terms involving $g_i g_j$ with $i\ne j$ in the RG evolution equations (\ref{SMEFTrge}), it is useful to redefine the operators in the Warsaw basis containing field-strength tensors by absorbing appropriate powers of the gauge couplings. This is particularly convenient, because in background-field gauge the combinations $g_s\spac G_\mu$, $g_2\spac W_\mu$ and $g_1 B_\mu$ are not renormalized \cite{Abbott:1980hw}. We denote the redefined operators by $Q_i'$ and their source terms by $S_i'$. All operators not listed here are defined as in the Warsaw basis.

\vspace{3mm}
\noindent
\underline{Class $X^3$:}\\[2mm]
We introduce new operators
\begin{equation}
\begin{aligned}
   Q_G' &= g_s\spac f^{abc}\spac G_\mu^{~\nu,a}\spac G_\nu^{~\rho,b}\spac G_\rho^{~\mu,c} \,, \\
   Q_W' &= g_2\spac\epsilon^{IJK}\spac W_\mu^{~\nu,I}\spac W_\nu^{~\rho,J}\spac W_\rho^{~\mu,K} \,,
\end{aligned}
\end{equation}
and similarly for the operators $Q_{\widetilde G}'$ and $Q_{\widetilde W}'$. The non-vanishing source terms for the new operators are given by
\begin{equation}
\begin{aligned}
   S_G' &= 8\spac C_{GG}^2 \,, \\
   S_W' &= 8\spac C_{WW}^2 \,.
\end{aligned}
\end{equation}

\vspace{3mm}
\noindent
\underline{Class $X^2 H^2$:}\\[2mm]
We introduce new operators
\begin{equation}
\begin{aligned}
   Q_{HG}' &= g_s^2\,H^\dagger H\,G_{\mu\nu}^a\spac G^{\mu\nu,a} \,, \\
   Q_{HW}' &= g_2^2\,H^\dagger H\,W_{\mu\nu}^I\spac W^{\mu\nu,I} \,, \\
   Q_{HB}' &= g_1^2\,H^\dagger H\,B_{\mu\nu}\spac B^{\mu\nu} \,, \\
   Q_{HW\negspac B}' &= g_1 g_2\,H^\dagger\sigma^I H\,W_{\mu\nu}^I\spac B^{\mu\nu} \,,
\end{aligned}
\end{equation}
and similarly for the operators $Q_{H\widetilde G}'$, $Q_{H\widetilde W}'$, $Q_{H\widetilde B}'$ and $Q_{H\widetilde W\negspac B}'$. The non-vanishing source terms for the new operators are given by
\begin{equation}
\begin{aligned}
   S_{HW}' &= - 2\spac C_{WW}^2 \,, \\
   S_{HB}' &= - 2\spac C_{BB}^2 \,, \\
   S_{HW\negspac B}' &= - 4\spac C_{WW}\spac C_{BB} \,.
\end{aligned}
\end{equation}

\vspace{3mm}
\noindent
\underline{Class $\psi^2 X H$:}\\[2mm]
We introduce new operators
\begin{equation}
\begin{aligned}
   Q_{eW}' &= g_2\spac\bar L_p\spac H\spac\sigma^{\mu\nu}\spac W_{\mu\nu}\,e_r \,, \\
   Q_{eB}' &= g_1\spac\bar L_p\spac H\spac\sigma^{\mu\nu} B_{\mu\nu}\,e_r \,, \\
   Q_{uG}' &= g_s\spac\bar Q_p\spac\tilde H\spac\sigma^{\mu\nu}\spac G_{\mu\nu}\,u_r \,, \\
   Q_{uW}' &= g_2\spac\bar Q_p\spac\tilde H\spac\sigma^{\mu\nu}\spac W_{\mu\nu}\,u_r \,, \\
   Q_{uB}' &= g_1\spac\bar Q_p\spac\tilde H\spac\sigma^{\mu\nu} B_{\mu\nu}\,u_r \,, \\
   Q_{dG}' &= g_s\spac\bar Q_p\spac H\spac\sigma^{\mu\nu}\spac G_{\mu\nu}\,d_r \,, \\
   Q_{dW}' &= g_2\spac\bar Q_p\spac H\spac\sigma^{\mu\nu}\spac W_{\mu\nu}\,d_r \,, \\
   Q_{dB}' &= g_1\spac\bar Q_p\spac H\spac\sigma^{\mu\nu} B_{\mu\nu}\,d_r \,, 
\end{aligned}
\end{equation}
where $W_{\mu\nu}=W_{\mu\nu}^I\spac\frac{\sigma^I}{2}$ and $G_{\mu\nu}=G_{\mu\nu}^a\spac T^a$. Their source terms are given by
\begin{equation}
\begin{aligned}
   \bm{S}_{eW}' &= - 2i\spac\widetilde{\bm{Y}}_e\,C_{WW} \,, \\
   \bm{S}_{eB}' &= - 2i \left( {\cal Y}_L + {\cal Y}_e \right) \widetilde{\bm{Y}}_e\,C_{BB} \,, \\
   \bm{S}_{uG}' &= - 4i\spac\widetilde{\bm{Y}}_u\,C_{GG} \,, \\
   \bm{S}_{uW}' &= - 2i\spac\widetilde{\bm{Y}}_u\,C_{WW} \,, \\
   \bm{S}_{uB}' &= - 2i \left( {\cal Y}_Q + {\cal Y}_u \right) \widetilde{\bm{Y}}_u\,C_{BB} \,, \\
   \bm{S}_{dG}' &= - 4i\spac\widetilde{\bm{Y}}_d\,C_{GG} \,, \\
   \bm{S}_{dW}' &= - 2i\spac\widetilde{\bm{Y}}_d\,C_{WW} \,, \\
   \bm{S}_{dB}' &= - 2i \left( {\cal Y}_Q + {\cal Y}_d \right) \widetilde{\bm{Y}}_d\,C_{BB} \,. 
\end{aligned}
\end{equation}

\section{\boldmath RG evolution of ALP couplings}
\label{app:B}
\renewcommand{\theequation}{B.\arabic{equation}}
\setcounter{equation}{0}

A detailed study of the RG evolution of the ALP couplings to SM particles in the effective ALP Lagrangians (\ref{Leff}) and (\ref{rela1}) has been performed in \cite{Bauer:2020jbp,Chala:2020wvs}. In order to meaningfully derive the one-loop approximations to the RG equations, one needs to make an assumption about the relative size of the ALP--boson and ALP--fermion couplings. We find it convenient to distinguish the scenarios of i) a ``natural ALP'', whose couplings respect the counting rules of naive dimensional analysis (see Section~\ref{sec:2}), and ii) a ``gauge-philic ALP'', whose couplings to gauge fields are parametrically enhanced by approximately a one-loop factor. Such an enhancement could arise, for example, from a parametrically large number $N_f\gg 1$ of new heavy fermions, by which the ALP couples to gauge fields. More specifically, we assume
\begin{equation}
\begin{aligned}
   &\text{Natural ALP:} &\qquad
    |c_{VV}| &\sim |\bm{c}_F| \,, &\qquad
    |C_{VV}| &\ll |\bm{c}_F| \,, \\
   &\text{Gauge-philic ALP:} &\quad
    |c_{VV}| &\gg |\bm{c}_F| \,, &\quad
    |C_{VV}| &\sim |\bm{c}_F| \,.
\end{aligned}
\end{equation}

\paragraph{Gauge-philic ALP.}

When translated to the notation used in (\ref{Leffalt}), and under the assumption that the ALP couplings $C_{VV}$ and $\widetilde{\bm{Y}}_f$ in this effective Lagrangian are of the same order, the one-loop evolution equations for the ALP--boson couplings derived in \cite{Bauer:2020jbp} take the form 
\begin{equation}
\begin{aligned}
   \frac{d}{d\ln\mu}\,C_{GG} 
   &= - \beta_0^{(3)}\,\frac{\alpha_s}{2\pi}\,C_{GG} \,, \\
   \frac{d}{d\ln\mu}\,C_{WW} 
   &= - \beta_0^{(2)}\,\frac{\alpha_2}{2\pi}\,C_{WW} \,, \\
   \frac{d}{d\ln\mu}\,C_{BB} 
   &= - \beta_0^{(1)}\,\frac{\alpha_1}{2\pi}\,C_{BB} \,,
\end{aligned}
\end{equation}
where $\beta_0^{(i)}$ are the one-loop coefficients of the $\beta$-functions of the three gauge groups, which above the electroweak scale are given by $\beta_0^{(3)}=7$, $\beta_0^{(2)}=\frac{19}{6}$ and $\beta_0^{(1)}=-\frac{41}{6}$. The one-loop evolution equations for the ALP--fermion couplings read
\begin{align}\label{eq:B.3}
   \frac{d}{d\ln\mu}\,\bm{\widetilde{Y}}_u
   &= \frac{1}{16\pi^2} \left[ 2\spac\widetilde{\bm{Y}}_u \bm{Y}_u^\dagger \bm{Y}_u
    + \frac52\,\bm{Y}_u \bm{Y}_u^\dagger \widetilde{\bm{Y}}_u
    - \frac32\,\bm{Y}_d \bm{Y}_d^\dagger \widetilde{\bm{Y}}_u
    - 2\spac\bm{Y}_d \widetilde{\bm{Y}}_d^\dagger \bm{Y}_u
    - \widetilde{\bm{Y}}_d \bm{Y}_d^\dagger \bm{Y}_u \right] \notag\\
   &\quad + i\spac\bm{Y}_u \left[ \frac{6\alpha_s}{\pi}\,C_F^{(3)}\,C_{GG}
    + \frac{3\alpha_2}{\pi}\,C_F^{(2)}\,C_{WW}
    + \frac{3\alpha_1}{\pi} \left( \mathcal{Y}_u^2 + \mathcal{Y}_Q^2 \right) C_{BB} \right] \notag\\
   &\quad - \widetilde{\bm{Y}}_u \left( \frac{3\alpha_s}{2\pi}\,C_F^{(3)} 
    + \frac{3\alpha_2}{4\pi}\,C_F^{(2)}
    + \frac{17\alpha_1}{48\pi} - \frac{T}{16\pi^2} \right) 
    - \bm{Y}_u\,\frac{\widetilde T}{8\pi^2} \,, \notag\\
   \frac{d}{d\ln\mu}\,\bm{\widetilde{Y}}_d 
   &= \frac{1}{16\pi^2} \left[ 2\spac\widetilde{\bm{Y}}_d \bm{Y}_d^\dagger \bm{Y}_d
    + \frac52\,\bm{Y}_d \bm{Y}_d^\dagger \widetilde{\bm{Y}}_d
    - \frac32\,\bm{Y}_u \bm{Y}_u^\dagger \widetilde{\bm{Y}}_d
    - 2\spac\bm{Y}_u \widetilde{\bm{Y}}_u^\dagger \bm{Y}_d
    - \widetilde{\bm{Y}}_u \bm{Y}_u^\dagger \bm{Y}_d \right] \notag\\
   &\quad + i\spac\bm{Y}_d \left[ \frac{6\alpha_s}{\pi}\,C_F^{(3)}\,C_{GG}
    + \frac{3\alpha_2}{\pi}\,C_F^{(2)}\,C_{WW}
    + \frac{3\alpha_1}{\pi} \left( \mathcal{Y}_d^2 + \mathcal{Y}_Q^2 \right) C_{BB} \right] \notag\\
   &\quad - \widetilde{\bm{Y}}_d \left( \frac{3\alpha_s}{2\pi}\,C_F^{(3)} 
    + \frac{3\alpha_2}{4\pi}\,C_F^{(2)} 
    + \frac{5\alpha_1}{48\pi} - \frac{T}{16\pi^2} \right) 
    + \bm{Y}_d\,\frac{\widetilde T}{8\pi^2} \,, \notag\\
   \frac{d}{d\ln\mu}\,\widetilde{\bm{Y}}_e
   &= \frac{1}{16\pi^2} \left[ 2\spac\widetilde{\bm{Y}}_e \bm{Y}_e^\dagger \bm{Y}_e
    + \frac52\,\bm{Y}_e \bm{Y}_e^\dagger \widetilde{\bm{Y}}_e \right] 
    + i\spac\bm{Y}_e \left[ \frac{3\alpha_2}{\pi}\,C_F^{(2)}\,C_{WW}
    + \frac{3\alpha_1}{\pi} \left( \mathcal{Y}_e^2 + \mathcal{Y}_L^2 \right) C_{BB} \right] \notag\\ 
   &\quad - \widetilde{\bm{Y}}_e \left( \frac{3\alpha_2}{4\pi}\,C_F^{(2)} 
    + \frac{15\alpha_1}{16\pi} - \frac{T}{16\pi^2} \right) + \bm{Y}_e\,\frac{\widetilde T}{8\pi^2} \,, 
\end{align}
where $C_F^{(N)}=\frac{N^2-1}{2N}$ denotes the eigenvalue of the quadratic Casimir operator in the fundamental representation of $SU(N)$, and
\begin{equation}
\begin{aligned}
   T &= \text{Tr}\Big[ N_c\spac\big( \bm{Y}_d^\dagger \bm{Y}_d + \bm{Y}_u^\dagger \bm{Y}_u \big)
    + \bm{Y}_e^\dagger \bm{Y}_e \Big] \,, \\
   \widetilde T &= \text{Tr} \left[ N_c\spac\big( \bm{Y}_d^\dagger\widetilde{\bm{Y}}_d 
    - \bm{Y}_u^\dagger\widetilde{\bm{Y}}_u \big) + \bm{Y}_e^\dagger\widetilde{\bm{Y}}_e \right] . 
\end{aligned}
\end{equation}
Note that we treat the SM Yukawa couplings as ${\cal O}(1)$ quantities, even though in practice their entries are highly hierarchical.

\paragraph{Natural ALP.}

In this case the ALP--boson couplings $C_{VV}$ are one-loop suppressed, and hence we prefer to work with the coupling parameters $\tilde c_{VV}$ defined in (\ref{eq:6}). Then the one-loop evolution equations for the ALP--boson couplings take the form 
\begin{align}
   \frac{d}{d\ln\mu}\,\tilde c_{GG}
   &= - \frac{i}{8\pi^2}\,\text{Tr}\big( \widetilde{\bm{Y}}_u\spac\bm{Y}_u^\dagger 
     + \widetilde{\bm{Y}}_d\spac\bm{Y}_d^\dagger \big) \,, \notag\\
   \frac{d}{d\ln\mu}\,\tilde c_{WW}
   &= - \frac{i}{32\pi^2}\,\text{Tr}\Big[ N_c\spac\big( \widetilde{\bm{Y}}_u\spac\bm{Y}_u^\dagger 
     + \widetilde{\bm{Y}}_d\spac\bm{Y}_d^\dagger \big) 
     + \widetilde{\bm{Y}}_e\spac\bm{Y}_e^\dagger \Big] \,, \\
   \frac{d}{d\ln\mu}\,\tilde c_{BB}
   &= - \frac{i}{8\pi^2}\,\text{Tr}\Big[ 
    N_c\spac\big( {\cal Y}_u^2 + {\cal Y}_Q^2 \big)\spac\widetilde{\bm{Y}}_u\spac\bm{Y}_u^\dagger 
    + N_c\spac\big( {\cal Y}_d^2 + {\cal Y}_Q^2 \big)\spac\widetilde{\bm{Y}}_d\spac\bm{Y}_d^\dagger 
    + \big( {\cal Y}_e^2 + {\cal Y}_L^2 \big)\spac\widetilde{\bm{Y}}_e\spac\bm{Y}_e^\dagger \Big] \,. \notag
\end{align}
The one-loop evolution equations for the ALP--fermion couplings read
\begin{align}
   \frac{d}{d\ln\mu}\,\bm{\widetilde{Y}}_u
   &= \frac{1}{16\pi^2} \left[ 2\spac\widetilde{\bm{Y}}_u \bm{Y}_u^\dagger \bm{Y}_u
    + \frac52\,\bm{Y}_u \bm{Y}_u^\dagger \widetilde{\bm{Y}}_u
    - \frac32\,\bm{Y}_d \bm{Y}_d^\dagger \widetilde{\bm{Y}}_u
    - 2\spac\bm{Y}_d \widetilde{\bm{Y}}_d^\dagger \bm{Y}_u
    - \widetilde{\bm{Y}}_d \bm{Y}_d^\dagger \bm{Y}_u \right] \notag\\
   &\quad - \widetilde{\bm{Y}}_u \left( \frac{3\alpha_s}{2\pi}\,C_F^{(3)} 
    + \frac{3\alpha_2}{4\pi}\,C_F^{(2)}
    + \frac{17\alpha_1}{48\pi} - \frac{T}{16\pi^2} \right) 
    - \bm{Y}_u\,\frac{\widetilde T}{8\pi^2} \,, \notag\\
   \frac{d}{d\ln\mu}\,\bm{\widetilde{Y}}_d 
   &= \frac{1}{16\pi^2} \left[ 2\spac\widetilde{\bm{Y}}_d \bm{Y}_d^\dagger \bm{Y}_d
    + \frac52\,\bm{Y}_d \bm{Y}_d^\dagger \widetilde{\bm{Y}}_d
    - \frac32\,\bm{Y}_u \bm{Y}_u^\dagger \widetilde{\bm{Y}}_d
    - 2\spac\bm{Y}_u \widetilde{\bm{Y}}_u^\dagger \bm{Y}_d
    - \widetilde{\bm{Y}}_u \bm{Y}_u^\dagger \bm{Y}_d \right] \notag\\
   &\quad - \widetilde{\bm{Y}}_d \left( \frac{3\alpha_s}{2\pi}\,C_F^{(3)} 
    + \frac{3\alpha_2}{4\pi}\,C_F^{(2)} 
    + \frac{5\alpha_1}{48\pi} - \frac{T}{16\pi^2} \right) 
    + \bm{Y}_d\,\frac{\widetilde T}{8\pi^2} \,, \notag\\
   \frac{d}{d\ln\mu}\,\widetilde{\bm{Y}}_e
   &= \frac{1}{16\pi^2} \left[ 2\spac\widetilde{\bm{Y}}_e \bm{Y}_e^\dagger \bm{Y}_e
    + \frac52\,\bm{Y}_e \bm{Y}_e^\dagger \widetilde{\bm{Y}}_e \right] \notag\\
   &\quad - \widetilde{\bm{Y}}_e \left( \frac{3\alpha_2}{4\pi}\,C_F^{(2)} 
    + \frac{15\alpha_1}{16\pi} - \frac{T}{16\pi^2} \right) + \bm{Y}_e\,\frac{\widetilde T}{8\pi^2} \,. 
\end{align}

\end{appendix}

\end{document}